\documentclass[twocolumn]{aastex631}
\usepackage{graphicx}
\usepackage[T1]{fontenc}
\usepackage{amssymb}
\usepackage{amsmath}
\usepackage{CJK}

\begin{document}
\begin{CJK*}{UTF8}{gbsn}

\title{Global Simulations of Gravitational Instability in Protostellar Disks with Full Radiation Transport\\II. Locality of Gravitoturbulence, Clumpy Spirals, and Implications for Observable Substructure}

\author[0000-0002-9408-2857]{Wenrui Xu (许文睿)}
\affiliation{Center for Computational Astrophysics, Flatiron Institute, New York, NY 10010, USA}
\author[0000-0002-2624-3399]{Yan-Fei Jiang (姜燕飞)}
\affiliation{Center for Computational Astrophysics, Flatiron Institute, New York, NY 10010, USA}
\author[0000-0003-1676-6126]{Matthew W. Kunz}
\affiliation{Department of Astrophysical Sciences, Princeton University, Peyton Hall, Princeton, NJ 08544, USA}
\affiliation{Princeton Plasma Physics Laboratory, PO Box 451, Princeton, NJ 08543, USA}
\author[0000-0001-5603-1832]{James M. Stone}
\affiliation{Institute for Advanced Study, 1 Einstein Drive, Princeton, NJ, 08540, USA}

\begin{abstract}
Spiral perturbations in a gravitationally unstable accretion disk regulate disk evolution through angular-momentum transport and heating and provide an observational signature of gravitational instability (GI).
We use global 3D simulations to systematically characterize and understand these spiral perturbations.
The spiral perturbations and the resulting transport are overall insensitive to the cooling type, with the exception that radiative cooling, especially in the optically thick regime, reduces the amplitude of temperature perturbations.
Spiral perturbations are localized around corotation, allowing transport to be approximated by a local $\alpha$ viscosity to zeroth order in aspect ratio ($H/R$), but only after averaging over multiple orbits in time and/or multiple scale heights in space. Meanwhile, large-amplitude perturbations from strong gravitoturbulence can cause $\mathcal O(\alpha^{1/2})$ deviation in the cooling rate of the disk.
We develop empirical prescriptions for the angular-momentum transport, heating, and cooling in a gravitoturbulent disk that capture the deviation from a viscous, unperturbed disk to first order in $H/R$ and $\alpha^{1/2}$.
The spiral perturbations in saturated gravitoturbulence are clumpy, with dense clumps forming through the nonlinear coupling between multiple modes at different $m$.
Observationally, the clumpy gravitoturbulence produced by saturated GI can be mistaken with observational noise or embedded companions, especially under finite resolution. Meanwhile, grand-design spirals with $m$-fold symmetry may be uncommon among disks in saturated gravitoturbulence, and we speculate that they may instead be a signature of recently triggered or decaying GI.
\end{abstract}

\section{Introduction}
Gravitational instability (GI) may occur in many astrophysical accretion disks, such as galactic disks, AGN disks, and protostellar disks. Specifically, in protostellar disks, GI can be an important ingredient of disk evolution and binary/planet formation. Gravitational fragmentation may allow direct formation of stellar/planetary companions. The perturbations associated with GI, which take the form of spirals/clumps/gravitoturbulence (there is often no clear distinction between these terms in the literature), regulate disk evolution and dust coagulation within the disk.

In \citet[][hereafter Paper I]{paper1} we reviewed theoretical and observational evidence for GI in protostellar disks, its role in disk evolution and planet formation, and a few outstanding issues with our theoretical understanding of GI. To address these issues, we perform and analyze a set of global 3D simulations (detailed in Paper I) surveying different cooling rates and thermodynamic prescriptions. We cover constant-$\beta$ and radiative cooling in both optically thin and optically thick regimes. In Paper I, we focused on issues related to fragmentation; in the current paper, we focus on issues related to spiral perturbations. 

This paper is motivated mainly by the following question: How should one model quantitatively the morphological properties of spirals and the transport they produce? This is a particularly timely question because recent observations of young protostellar disks \citep[e.g.,][]{Huang+18,Tobin+20,Aso+21} have enabled quantitative comparison with self-consistent models of disks that are self-regulated by GI \citep[e.g.,][]{Xu22,Xu+23}. However, existing prescriptions for modeling spiral substructures and transport exhibit several limitations and inconsistencies.

When modeling the transport associated with GI, a common assumption is that the angular-momentum transport and heating produced by GI is equivalent to that of a local viscosity \citep{Gammie01}. This assumption is correct in the local (shearing-box) limit, but in a global setup, self-gravity leads to nonlocal transport whenever spiral perturbations are not localized around corotation \citep{BalbusPapaloizou99}. Consistent with these theoretical arguments, simulations have demonstrated an empirical trend that GI can be modeled as a local viscosity for a low disk-to-star mass ratio \citep[e.g.,][]{LodatoRice04,Bethune+21}, but not for more massive disks which tend to exhibit more global spiral perturbations \citep[e.g.,][]{LodatoRice05,Forgan+11}. However, we lack a more quantitative perscription for estimating how much the transport deviates from a local viscosity. It also remains unclear whether factors other than the disk-to-star mass ratio, such as the amplitude of the perturbation, are relevant.

Another major limitation of this picture is that it does not address how GI affects the cooling rate of the disk. Estimating the rate of radiative cooling as a function of basic disk properties such as average surface density and average temperature is important for constructing semi-analytic disk models used in observational modeling and planet formation calculations. Although the cooling rate of an unperturbed disk can be accurately estimated (e.g., Appendix A of Paper I), it remains unclear whether large-amplitude perturbations driven by GI can significantly affect the efficiency of radiative cooling.

When modeling spiral substructures driven by GI, there is often a disconnect between theory and simulation. The theoretical foundation for modeling spirals in gravitationally unstable disks is laid out by \citet{LinShu66}. They assume that the perturbations are dominated by a single linear eigenmode, or a single mode plus its overtones when nonlinear saturation is considered (see a review in \citealt{Shu16}). This allows spiral properties, such as pitch angle, to be solved from the linear dispersion relation. A distinguishing feature of the Lin \& Shu picture is the presence of global ``grand-design'' spirals that show exact $m$-fold symmetry. However, simulations exhibit a continuum of complex morphologies (for example, see Fig.~1 of \citealt{Forgan+11}) that do not conform to this picture. Some examples from this continuum include global spirals in thick and massive disks; localized gravitoturbulence in thin and less massive disks; and dense clumps in disks with strong GI on the verge of fragmentation. Among these examples, only the first qualitatively resembles the Lin \& Shu picture; even in that case, simulations frequently show spirals that deviate significantly from exact $m$-fold symmetry \citep[e.g.,][]{LodatoRice05,Forgan+11}. In this paper, we will explore how the Lin \& Shu picture could potentially be extended to cover all these regimes by focusing on the coexistence and coupling between modes with different $m$.

Developing a more comprehensive theoretical framework for spiral perturbations is also crucial for interpreting observations. Among the protoplanetary disks that show evidence of GI, some have spirals that visually resemble the symetric grand-design spirals in the Lin \& Shu picture; some examples include Elias 2-27 \citep{Paneque-Carreno+21} and TMC1A \citep{Xu+23}. Others, however, have spirals that deviate significantly from $m$-fold symmetry and more closely resemble the ones seen in simulations; some examples include AB~Aur \citep{Speedie+24} and RU~Lup \citep{Huang+20}. In this paper, we will also discuss briefly how the deviation from a classic Lin \& Shu spiral could affect the observational signatures of GI.

The remainder of this paper is organized as follows.
In Sections~\ref{sec:setup} and \ref{sec:overview} we summarize our simulation setup and provide an overview of simulation results.
In Section~\ref{sec:transport} we discuss the transport, heating, and cooling produced by spiral perturbations.
In Section~\ref{sec:spiral} and Section~\ref{sec:spiral:obs} we discuss the physical picture of spiral perturbations and its observational implications.
We summarize our findings in Section~\ref{sec:conclusion}.

\section{Simulation setup}\label{sec:setup}
We perform global hydrodynamic and radiation hydrodynamic simulations of a self-gravitating disk around a central point mass (star) using the \texttt{Athena++} code suite \citep{Stone+20}. This section summarizes our simulation setup; for a more detailed description, see Section~2 of Paper~I.

\subsection{Units, coordinates, and resolution}
We choose the unit mass $m_0$ to be the total (combined) mass of the star and the disk $M_{\rm tot}$, the unit length $r_0$ to be the initial disk size, and the unit time to be $t_0 \equiv \Omega_0^{-1} \equiv (GM_{\rm tot}/r_0^3)^{1/2}$. The simulations adopt a spherical-polar domain, though some of our definitions and diagnostics require cylindrical coordinates; we use $(r,\theta,\phi)$ and $(R,\phi,z)$ to denote spherical and cylindrical coordinates, respectively. We adopt a non-uniform grid, and the resolution near the midplane corresponds to ${\approx}(1/12, 1/16, 1/4)$ scale heights in $(r,\theta,\phi)$. We also perform low-resolution runs with half of our fiducial resolution in all directions. For most results in this paper, simulations performed at low and fiducial resolutions agree quantitatively with each other.

\subsection{Initial condition}
We use a weakly gravitoturbulent disk as the initial condition for all simulations. To construct this initial condition, we first initialize a Keplerian torus rotating at
\begin{equation}
    \Omega_{\rm K} \equiv \sqrt{\frac{GM_{\rm tot}}{R^3}}.
\end{equation}
The torus spans $[r_0/3, r_0]$ with initial surface density and temperature profiles given by
\begin{equation}
    \Sigma_{\rm init} \propto R^{-2},~~~T_{\rm init} \propto R^{-1}.
\end{equation}
These profiles are motivated by the power-law slopes of gravitationally self-regulated disks seen in radiative simulations, semi-analytic modeling, and observations \citep[e.g.,][]{XK21b,Xu22,Xu+23}.
This initial condition corresponds to uniform aspect ratio $H/R$ and uniform ``Keplerian Toomre $Q$'', $Q_{\rm K}$. The scale height $H$ and $Q_{\rm K}$ are defined by
\begin{equation}
    H\equiv \frac{c_{\rm s,iso}}{R\Omega_{\rm K}},~~~Q_{\rm K} \equiv \frac{c_{\rm s}\Omega_{\rm K}}{\pi G \Sigma}.
\end{equation}
Here $c_{\rm s}$ is the sound speed and $c_{\rm s,iso}\equiv\gamma^{-1/2}c_{\rm s}$ is the isothermal sound speed; we adopt an adiabatic index of $\gamma=5/3$.
We choose an initial disk mass $M_{\rm d}=0.1M_{\rm tot}$ and initial $Q_{\rm K}=1.5$; this corresponds to $H/R=0.053$.
The torus is then evolved under weak cooling for $30 t_0$ to allow a self-consistent gravitoturbulent state to develop. This gravitoturbulent state is used as the initial condition for all of our simulations.

\subsection{Cooling and radiation}
Our simulations cover three different cooling types, outlined below.

\textbf{Constant $\beta$ cooling:} The normalized cooling time $\beta\equiv \Omega_{\rm K}t_{\rm cool}$ is constant throughout the disk. This prescription is not realistic, but is commonly adopted \citep[e.g.,][]{Gammie01} due to its simplicity.

\textbf{Optically thin cooling:} When the disk optical depth is negligible, the cooling rate per mass is ${\propto}\kappa T^4$. As in Paper~I, we assume that the opacity $\kappa$ is constant for each simulation. To survey different cooling rates, we tune the $\beta$ of our torus initial condition at $r_0$, denoted by $\beta_0$.

\textbf{Optically thick cooling:} This is similar to optically thin cooling, except that we also specify the initial optical depth at $r_0$, denoted by $\tau_0$. To capture the complex radiation dynamics at finite optical depth (especially at $\tau\sim 1$, where most cooling takes place), we adopt the implicit full radiation transport algorithm in \citet{Jiang21}. We still assume constant $\kappa$, and we ignore the scattering opacity for simplicity. Given the high cost of these simulations, we fix $\tau_0 =10$ and only vary $\beta_0$.

\section{Overview of simulation results}\label{sec:overview}
\begin{figure*}
    \centering
    \includegraphics[scale=0.66]{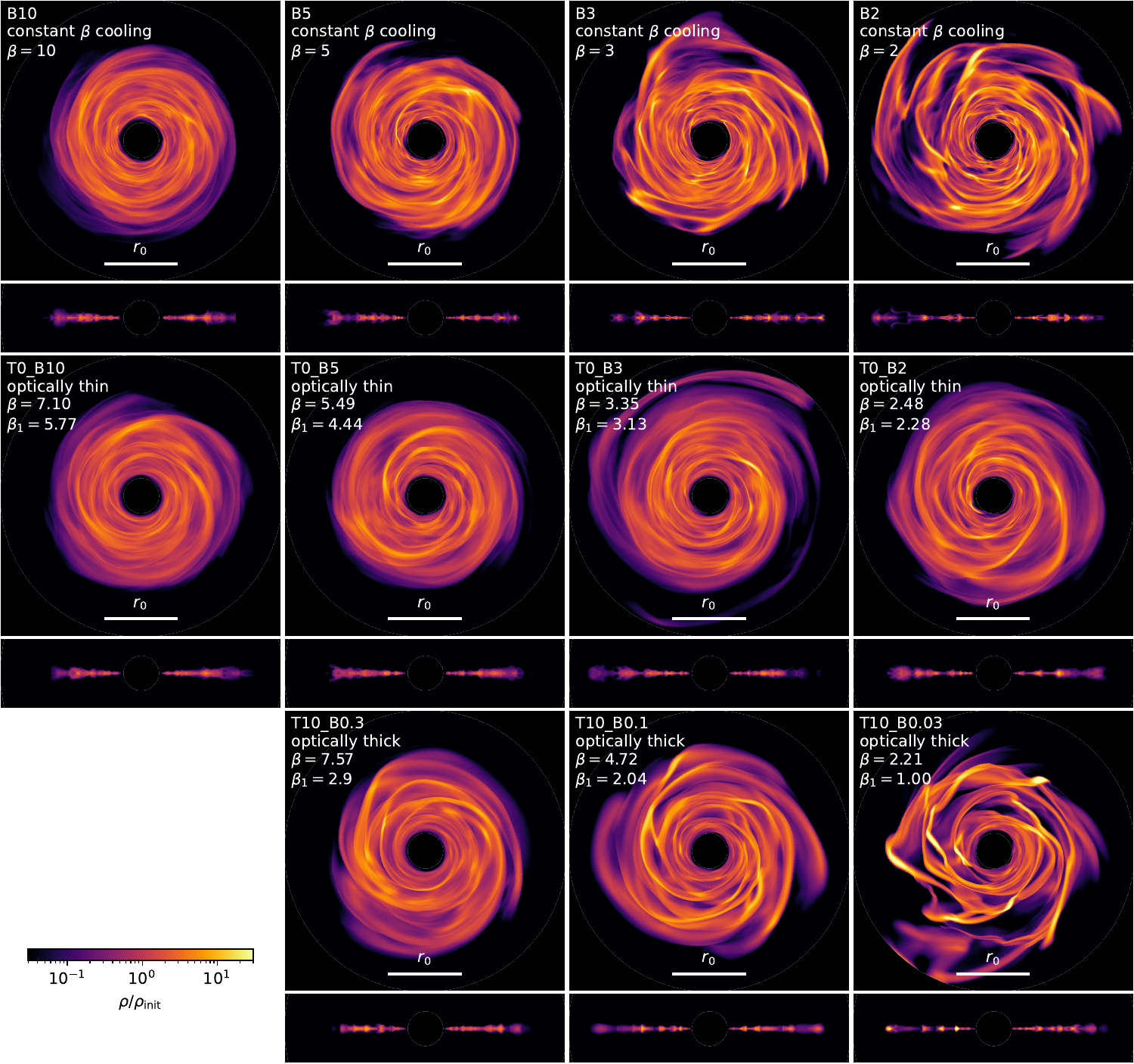}
    \caption{Density snapshots of fiducial-resolution simulations. We normalize density $\rho$ by the midplane density of our initial condition $\rho_{\rm init}(R)$. Each row shows a different cooling type. For each run, we also label the disk-averaged cooling time $\beta$ and the cooling time at $1 r_0$, $\beta_1$; both are averaged across the whole simulation.}
    \label{fig:overview_rho}
\end{figure*}
\begin{figure*}
    \centering
    \includegraphics[scale=0.66]{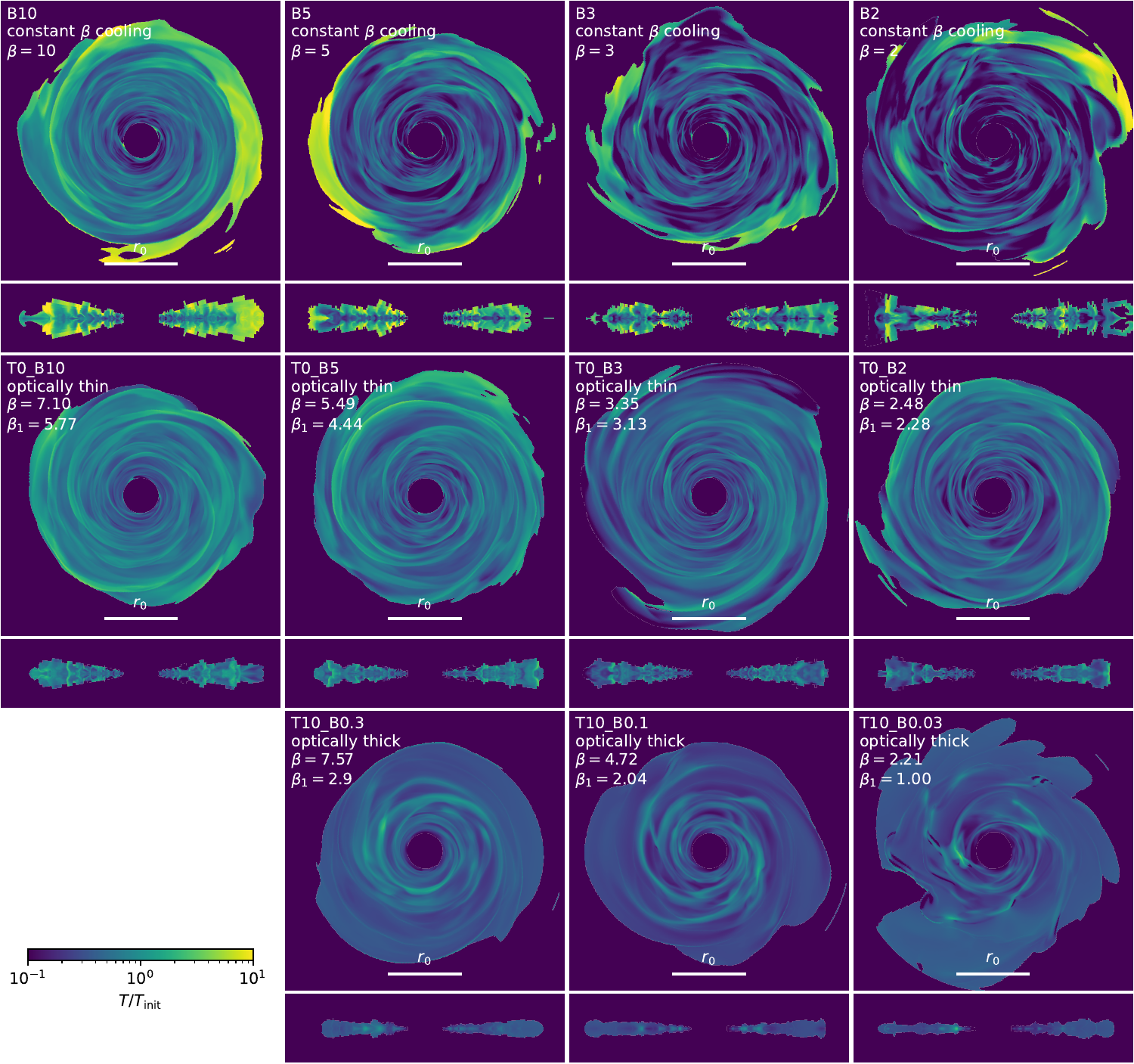}
    \caption{Similar to Fig.~\ref{fig:overview_rho}, but shows disk temperature normalized by the temperature of our initial condition $T_{\rm init}(R)$. Here we only plot cells with $\rho>10^{-4}\rho_0$. Radiative cooling reduces the amplitude of temperature perturbation, especially in the optically thick regime.}
    \label{fig:overview_T}
\end{figure*}

\begin{figure*}
    \centering
    \includegraphics[scale=0.66]{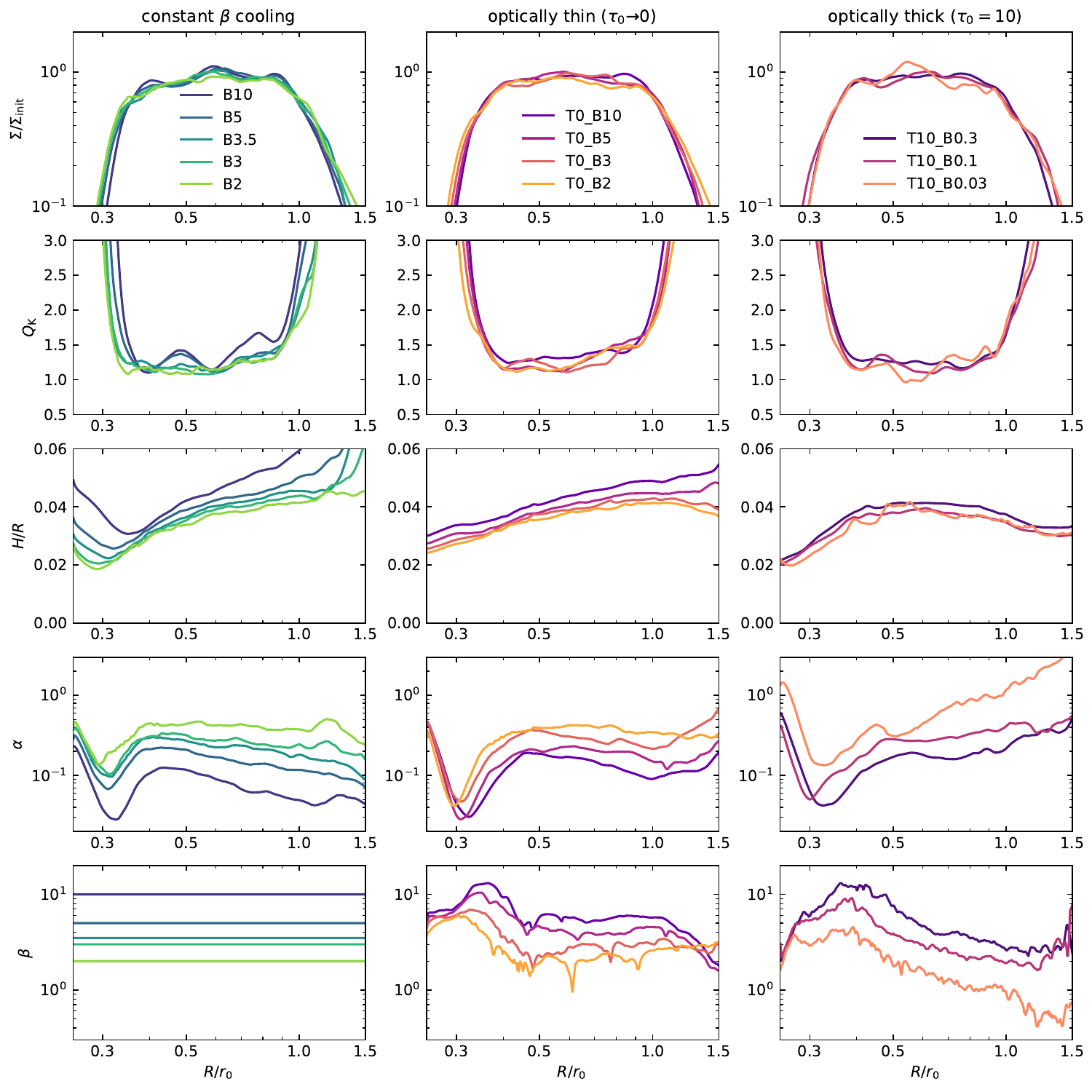}
    \caption{Time averaged radial profiles. From top to bottom, we show the surface density $\Sigma$ normalized by the initial density profile $\Sigma_{\rm init}$; the Keplerian Toomre $Q$, $Q_{\rm K}$; aspect ratio $H/R$; effective viscosity (or normalized angular-momentum transport rate) $\alpha$; and normalized cooling time $\beta$. The three columns show constant $\beta$ cooling, optically thin cooling, and optically thick cooling, respectively.}
    \label{fig:radial_profile}
\end{figure*}

For each cooling type and resolution, we perform a few simulations to survey different cooling rates ($\beta$ or $\beta_0$). Figs.~\ref{fig:overview_rho} and~\ref{fig:overview_T} show density and temperature snapshots from our fiducial-resolution simulations. We label the simulations by B$\beta$ for constant-$\beta$ runs and T$\tau_0$\_B$\beta_0$ for radiative cooling runs ($\tau_0=0$ denotes optically thin cooling); a full list of simulations is given in Table~1 of Paper~I. Most of our simulations run for $30t_0$.
For all simulations, we see complex spiral dynamics and there is no clear global modes with $m$-fold symmetry as commonly assumed in spiral density wave theory. We discuss the nature of these spiral perturbations in Section~\ref{sec:spiral}.
Comparing across snapshots at different cooling rates, we see a clear correlation between perturbation amplitude and cooling. Comparing across different cooling types, radiative runs produce smaller-amplitude temperature perturbations than constant-$\beta$ runs due to the steeper temperature dependence of the cooling rate. Additionally, optically thick runs produce smaller-amplitude temperature perturbations than optically thin runs. This is because, in the optically thick regime, radiation acts as an effective thermal conduction that smooths out temperature perturbations.
Except for this difference in temperature perturbations, at a given cooling rate the spiral properties show little difference between different cooling types, as we show in later sections.

Fig.~\ref{fig:radial_profile} summarizes the radial profiles of time-averaged properties from our fiducial-resolution simulations.
The surface density stays close to the initial condition ($\Sigma_{\rm init}$) since the disk's evolution time is short compared to the timescale of mass transport. The $Q_{\rm K}$ parameter is generally ${\approx}1$--$1.5$ in the gravitationally unstable region (i.e., where $Q_{\rm K}\lesssim 2$). Here $Q_{\rm K}$ is a good approximation of the Toomre~$Q$ parameter because the disk mass is small relative to the central point mass and the rotation profile is close to Keplerian. The aspect ratio $H/R$ is typically ${\approx}0.04$ in the gravitationally unstable region, close to the initial value of $H/R= 0.053$. Outside of the gravitationally unstable region, where the disk temperature is no longer set by the thermal saturation of GI, $H/R$ shows more variation depending on the details of local heating and cooling mechanisms. The last two panels of Fig.~\ref{fig:radial_profile} show the effective viscosity $\alpha$, which corresponds to the rate of angular-momentum transport, and the cooling timescale $\beta$, defined as\footnote{The definition of effective viscosity $\alpha$ can differ slightly across studies; another common choice is replacing the denominator $\langle p\rangle$ in Eq.~\eqref{eq:alpha_def} by $\frac 32 \langle\rho c_s^2\rangle$.}
\begin{equation}\label{eq:alpha_def}
    \alpha \equiv \frac{\langle S_{\rm h}+S_{\rm g} \rangle}{\langle p \rangle},~~~\beta \equiv \frac{\Omega_{\rm K}\langle u\rangle}{\langle -q \rangle}.
\end{equation}
Here $\langle\cdot\rangle$ denotes averaging in time on the spherical shell at $r=R$; $p$ is the pressure, $u$ the internal energy density, $-q$ the cooling rate per volume, and $S_{\rm h}$ and $S_{\rm g}$ the hydrodynamic and gravitational stresses defined by
\begin{equation}
    S_{\rm h} \equiv \rho v_r' v_\phi',~~~S_{\rm g}\equiv \frac{1}{4\pi G}g_r g_\phi.
\end{equation}
Here, $v_r'$ and $v_\phi'$ are the velocity perturbations with respect to the mean (density weighted, spherically averaged) velocity.
We note that our disks have relatively small $H/R$, so there is no significant difference between computing these properties on spherical and cylindrical shells; here we calculate them on spherical shells to align with our simulation grid.
In the gravitationally unstable region, $\alpha$ is anti-correlated with $\beta$. The $\alpha$ profiles often show some structure near the inner boundary, but that is mainly due to dynamics at the disk edge and in the low-density region near the inner boundary. We limit our attention to the gravitationally unstable region, and we discuss the transport, heating, and cooling there in the next section.

\section{Transport, heating, and cooling by GI}\label{sec:transport}
\subsection{Theoretical foundation for modeling GI as a local viscosity}\label{sec:transport:visc}

\begin{figure*}
    \centering
    \includegraphics[width=\textwidth]{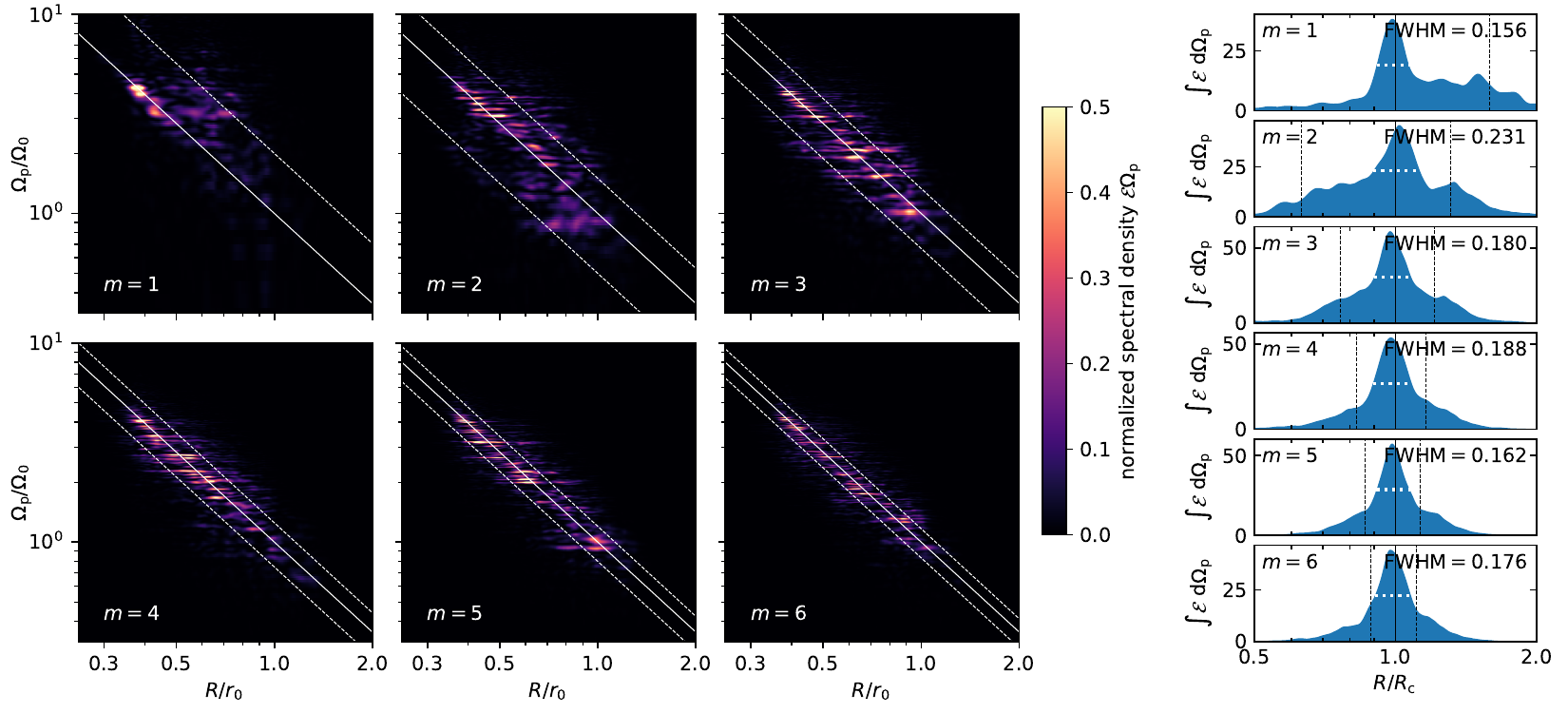}
    \caption{Left: Periodogram of density perturbations for run T10\_B0.1. Other runs show qualitatively similar results. The spectral density of surface density perturbation $\mathcal{E}(m,R,\Omega_{\rm p})$ is normalized such that $\sum_m\int\mathcal{E}\,{\rm d}\Omega_{\rm p} = \overline{\Sigma^2}(R)/\Sigma_{\rm init}^2(R)$. Here we plot $\mathcal{E}\Omega_{\rm p}$ which represents the contribution to $\overline{\Sigma^2}(R)/\Sigma_{\rm init}^2(R)$ per log frequency. Corotation and Lindblad resonances are marked by solid and dashed lines, respectively; these are estimated assuming Keplerian rotation, $\Omega = \kappa = \Omega_{\rm K}$. Each mode corresponds to a horizontal line in the diagram. Most modes are centered at corotation with a narrow radial span. Right: integrating the spectral density along lines of constant $R/R_{\rm c}$ ($R_{\rm c}$ is the location of corotation resonance) to quantify the typical radial extent of each mode. The FWHM of the distribution, which serves as a measure of the typical radial extent of modes, is marked by white dotted lines. Modes can reach a distance comparable to the location of the Lindblad resonances, but the FWHM remains only ${\approx}0.2$ or ${\approx}5$ scale heights.}
    \label{fig:periodogram}
\end{figure*}

\begin{figure*}
    \centering
    \includegraphics[scale=0.66]{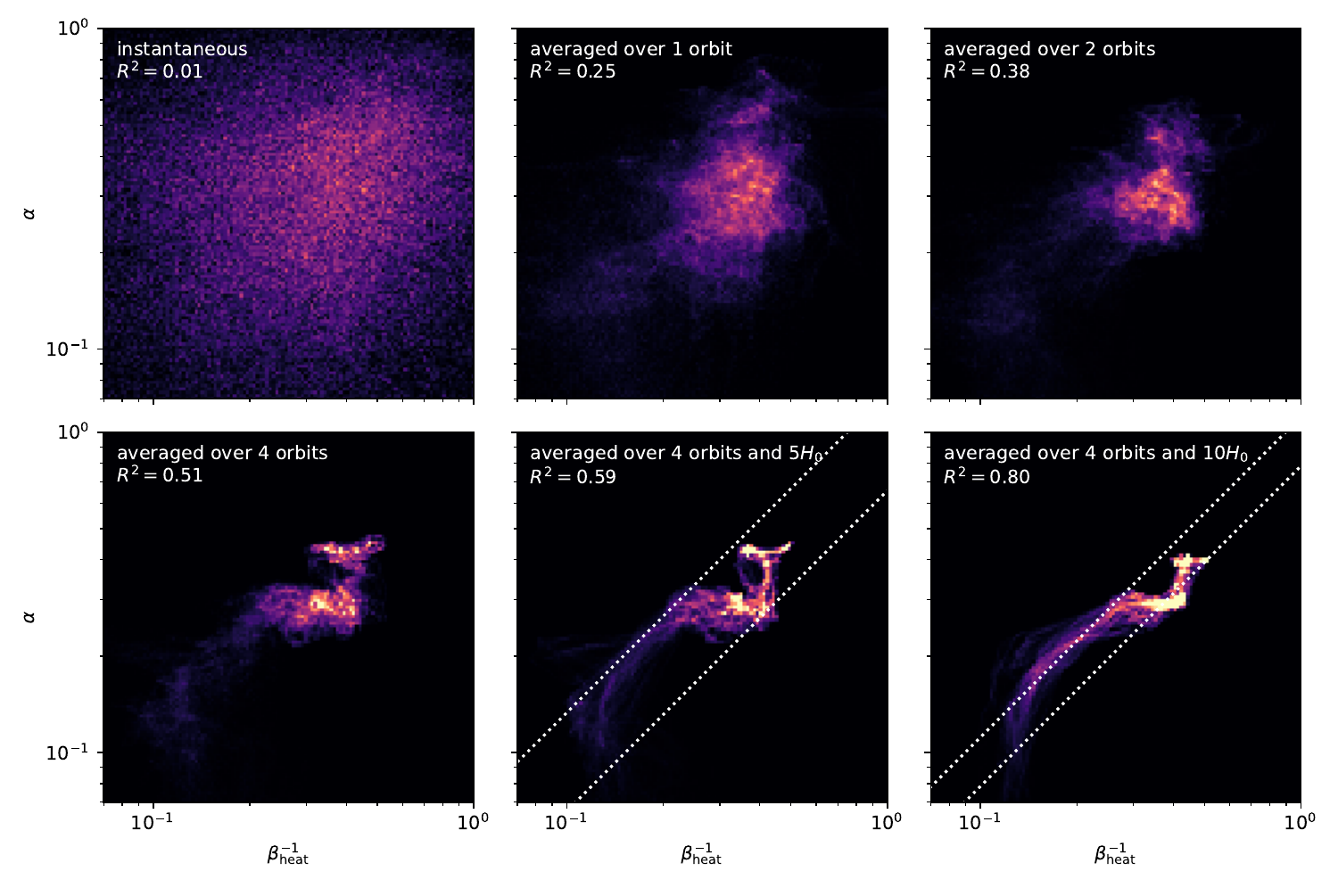}
    \caption{The relation between the heating rate $\beta_{\rm heat}^{-1}$ and the effective $\alpha$ in run T10\_B0.1, under different levels of averaging. Color marks the mass-weighted probability density. We also label the $R^2$ value when fitting the data to a linear relation between $\log\alpha$ and $\log\beta_{\rm heat}^{-1}$. The averaging is performed with a moving window corresponding to a certain number of local orbital times $2\pi\Omega_{\rm K}^{-1}$ and a certain number of typical scale heights $H_0\equiv 0.04 R$. Although the instantaneous result shows a high level of variability and no clear correlation, averaging produces an approximately (though not exactly) linear relation between $\alpha$ and $\beta_{\rm heat}^{-1}$. In the last two panels, we mark the estimated deviation (white dashed line, showing $\pm 1\sigma$) from a linear $\alpha$--$\beta_{\rm heat}^{-1}$ relation following Eq.~\eqref{eq:fluctuation_amplitude}.}
    \label{fig:variability}
\end{figure*}

\begin{figure}
    \centering
    \includegraphics[scale=0.66]{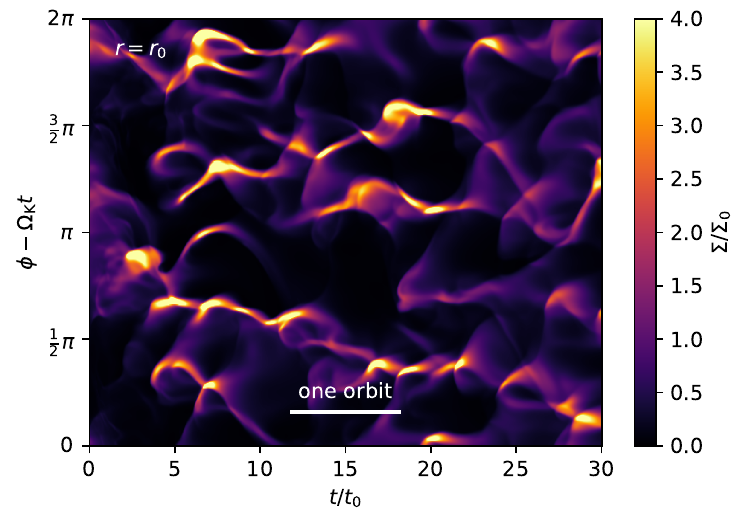}
    \includegraphics[scale=0.66]{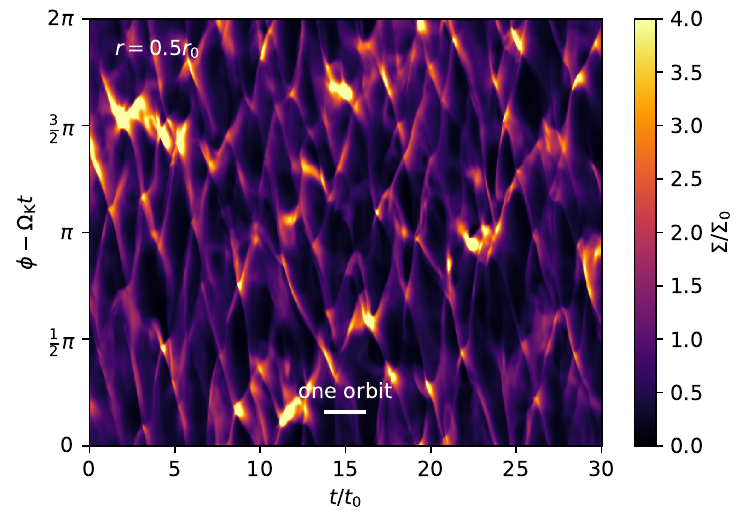}
    \caption{Time evolution of the surface density profile at given radii (top panel: $r_0$; bottom panel: $0.5 r_0$) for run T10\_B0.1. We apply an azimuthal shift of $\Omega_{\rm K} t$ to compensate for rotation. The horizontal span of a mode shows its lifetime, which is usually comparable to the orbital timescale (marked by white segments at the bottom of the panels).}
    \label{fig:spiral_lifetime}
\end{figure}

In this subsection, we review the theoretical foundation of modeling GI as a local viscosity and quantitatively demonstrate that it is a reasonable approximation for the geometrically thin disks in our simulations, consistent with previous studies \citep[e.g.,][]{LodatoRice04,Bethune+21}.
We also provide a new theoretical argument for the necessity of averaging when approximating GI as a local viscosity.

Modeling turbulent transport by an effective viscosity is motivated by a parallel between viscous stress and turbulent stress \citet{BalbusPapaloizou99}. For a local viscosity, the viscous angular-momentum flux, energy flux, and heating rate are all directly proportional to the viscous stress. To the lowest order in $\delta v/v_{\rm K}$, these properties are exactly preserved for hydrodynamic and magnetic-field perturbations if one were to replace the viscous stress by the appropriate turbulent stress. This holds even when the flow is not in a steady state. However, this relation no longer holds for self-gravity. With self-gravity, in addition to the viscous-like terms, there is an ``anomalous'' energy flux $F^*$ (\citealt{BalbusPapaloizou99}, Eq.~64) of order $\Omega_{\rm p}/\Omega_{\rm K}-1$, where $\Omega_{\rm p}$ is the pattern speed of the perturbation. In other words, the deviation from a local viscosity depends on how much the perturbations deviate from corotation.

Are the perturbations seen in our simulationsm close to corotation? To quantitatively answer this question, we decompose the surface density perturbation into modes with $\delta\Sigma\propto {\rm e}^{{\rm i}m\Omega_{\rm p} t-{\rm i}m\phi}$, where $m$ is the azimuthal wavenumber and $\Omega_{\rm p}$ is the pattern speed.
This decomposition produces the periodogram in Fig.~\ref{fig:periodogram}. Here, each mode ($m,\Omega_{\rm p}$) corresponds to a horizontal line in one of the panels. Modes appear throughout the disk, but each mode only spans a narrow range of radii around the corotation resonance $R_{\rm c}$ defined by $\Omega_{\rm p} = \Omega_{\rm K}(R_{\rm c})$. In the rightmost panel of Fig.~\ref{fig:periodogram}, we integrate the periodogram along $R\propto R_{\rm c}$ to quantify the typical radial spread of the modes. Most of the mode energy is concentrated near corotation with a FWHM of $\Delta R/R_{\rm c}\approx 0.2$, which corresponds to only ${\approx}2.5$ scale heights on each side of the resonance, similar to the findings in \citet{LodatoRice04}, \citet{Bethune+21}, and \citet{Steiman-Cameron+23}. The localization around corotation allows the perturbations to be approximated as a local viscosity.
Meanwhile, a small amount of perturbation can reach larger distances from corotation, with the maximum distance a mode can reach being comparable to the location of Lindblad resonances, similar to the finding in \citet{XK21b} and \citet{Steiman-Cameron+23}; the low-amplitude perturbation far from corotation does not play a significant role in energy and angular-momentum transport, but it can give the spirals a more global look, particularly in a log-scale plot like Fig.~\ref{fig:overview_rho}.

It is also worth noting that the small spread around corotation we observe is related to the small aspect ratio of the disks in our simulations. Since the scale height sets a typical radial scale for perturbations, in a thicker disk the perturbations could span a larger radial extent, resulting in larger deviation from local viscosity. This could explain the deviation seen in the literature that a higher disk mass, which for a marginally gravitationally unstable disk translates to a larger aspect ratio, produces a larger deviation from local viscosity \citep{LodatoRice05, Forgan+11}.

Another important but often overlooked implication of the anomalous flux argument of \citet{BalbusPapaloizou99} is that, in the presence of self-gravity, sufficient averaging in time and/or space is required for the local viscosity approximation to be valid, even when perturbations are highly localized around corotation. This is because the divergence of this anomalous energy flux $F^*$ can dominate instantaneous heating. Let $\lambda$ be the radial scale of the perturbation, which is also the scale of radial correlation. The amplitude of the stochastic heating associated with GI is
\begin{equation}
    \delta(\beta_{\rm heat}^*)^{-1} \sim \frac{1}{\Omega_{\rm K} U}\frac{\delta F^*}{\lambda} \sim \frac{1}{\Omega_{\rm K} U}\frac{F_0(\Omega_{\rm p}/\Omega_{\rm K}-1)}{\lambda} \sim \alpha.\label{eq:delta_beta_star}
\end{equation}
Here $F_0\equiv R\Omega_{\rm K} \alpha P$ is the typical energy flux, and $P$ and $U$ are the vertically integrated pressure and internal energy, respectively. As an order-magnitude estimate, the fluctuation amplitude is $\delta F^* \sim F^* \sim F_0|\Omega_{\rm p}/\Omega_{\rm K}-1|$. We note that $|\Omega_{\rm p}/\Omega_{\rm K}-1| \sim \lambda/R$ for corotating modes. Eq.~\eqref{eq:delta_beta_star} suggests that $F^*$ causes the instantaneous $\alpha$--$\beta_{\rm heat}$ relation to deviate from the viscous relation by order unity.
However, once we average out the fluctuations by averaging in time and radius, we are left with a very different result. If we only consider the effect of the mean $F^*$, the resulting heating is
\begin{equation}
    \overline{(\beta_{\rm heat}^*)^{-1}} \sim \frac{1}{\Omega_{\rm K} U}\frac{\overline{F^*}}{R} \sim \frac{\lambda}{R}\alpha\sim \frac{H}{R}\alpha.\label{eq:mean_beta_star}
\end{equation}
Here we have assumed $\lambda\propto H$ following Section~\ref{sec:spiral}.
The main difference between Eq.~\eqref{eq:delta_beta_star} and Eq.~\eqref{eq:mean_beta_star} is that the mean $F^*$ varies on a much more global scale (${\sim}R$) instead of $\lambda$, provided that the averaged disk properties do not contain small-scale radial substructures.\footnote{GI does have the potential to produce radial substructures, but those tend to be associated with Lindblad resonances and the length scale of radial variation is ${\sim}R$ (as opposed to ${\ll} R$); see, for example, \citet{XK21b} and \citet{Steiman-Cameron+23}.} Eq.~\eqref{eq:mean_beta_star} suggests that $F^*$ causes the averaged $\alpha$--$\beta_{\rm heat}^{-1}$ relation to deviate from the viscous relation only by $\mathcal O(H/R)$.

To test this argument with our simulations, in Fig.~\ref{fig:variability} we show the distribution of the normalized rate of angular-momentum transport $\alpha$ and the heating rate $\beta_{\rm heat}^{-1}$ for a simulation, under different levels of temporal and spatial averaging. Here $\beta_{\rm heat}^{-1}$ is calculated by first measuring the rate of change of the internal energy and then removing the contributions from radiative heating/cooling and advection/compression by the density-weighted mean radial velocity. The local and instantaneous distributions of $\alpha$ and $\beta_{\rm heat}^{-1}$ are very broad and barely show any correlation (top-left panel). However, after averaging across a few orbits, a correlation between $\alpha$ and $\beta_{\rm heat}^{-1}$ emerges; the correlation improves if we further average across a few scale heights. After averaging, the relation between $\alpha$ and $\beta_{\rm heat}^{-1}$ is approximately linear (bottom-right panel).

We can also estimate how the amplitude of fluctuation in the $\alpha$--$\beta_{\rm heat}^{-1}$ relation scales with the level of time and radial averaging. Consider the logarithmic deviation $\sigma$ from a linear $\alpha$--$\beta_{\rm heat}^{-1}$ relation due to finite duration of averaging; the instantaneous deviation is $\sigma\sim 1$, as we estimated in Eq.~\eqref{eq:delta_beta_star}. After averaging in space across a length scale $\lambda_{\rm avg}$ and in time across a timescale $t_{\rm avg}$, the deviation then becomes
\begin{equation}
    \sigma \sim \frac{1}{(t_{\rm avg}/t_{\rm corr})^{1/2}(\lambda_{\rm avg}/\lambda_{\rm corr})}.\label{eq:fluctuation_amplitude}
\end{equation}
Here the correlation timescale $t_{\rm corr}$ and the correlation length scale $\lambda_{\rm corr}$ can be estimated based on the typical lifetime and radial extent of clumps, giving $t_{\rm corr}\sim 2\pi\Omega_{\rm K}^{-1}$ (Fig.~\ref{fig:spiral_lifetime}) and $\lambda_{\rm corr}\sim 5H$ (Fig.~\ref{fig:periodogram}). In Eq.~\eqref{eq:fluctuation_amplitude}, the $(t_{\rm avg}/t_{\rm corr})^{1/2}$ term accounts for the time averaging of the anomalous energy flux $\delta F^*$, and $\lambda_{\rm avg}/\lambda_{\rm corr}$ accounts for the amount of disk mass affected by the $\delta F^*$ difference between the two ends of the averaged region. Note that this estimate assumes $t_{\rm avg}\gtrsim t_{\rm corr}$ and $\lambda_{\rm avg}\gtrsim \lambda_{\rm corr}$.
In the last two panels of Fig.~\ref{fig:variability}, we plot the estimated $\pm 1\sigma$ deviation from a linear $\alpha$--$\beta_{\rm heat}^{-1}$ relation following Eq.~\eqref{eq:fluctuation_amplitude}. The estimated $\sigma$ is consistent with the variation of the data.

In summary, the radially localized perturbation of GI allows it to be approximated by a local viscosity (up to an error of order $H/R$) in a time and/or spatially averaged sense, but not in an instantaneous sense.

\subsection{Modeling the effects of GI}

\begin{figure}
    \centering
    \includegraphics[scale=0.66]{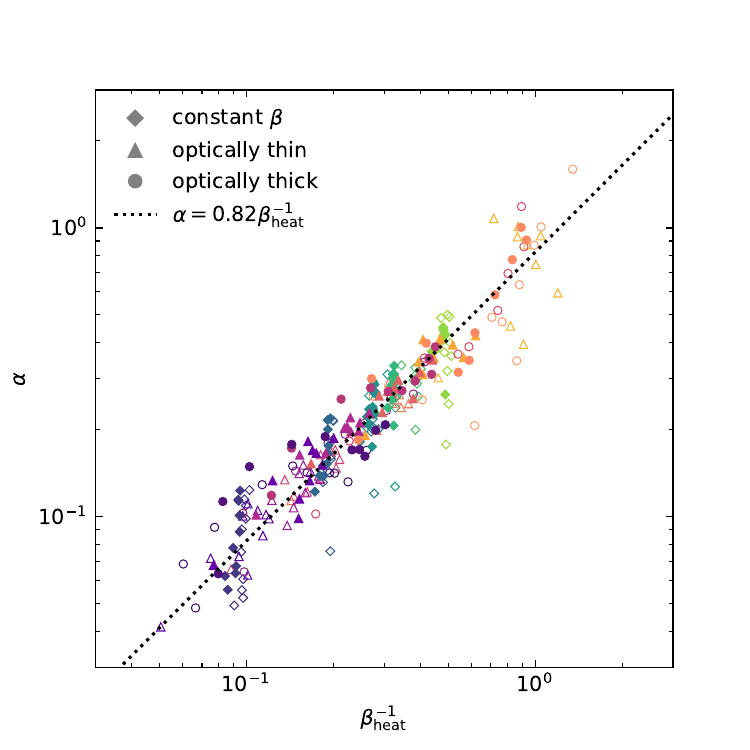}
    \includegraphics[scale=0.66]{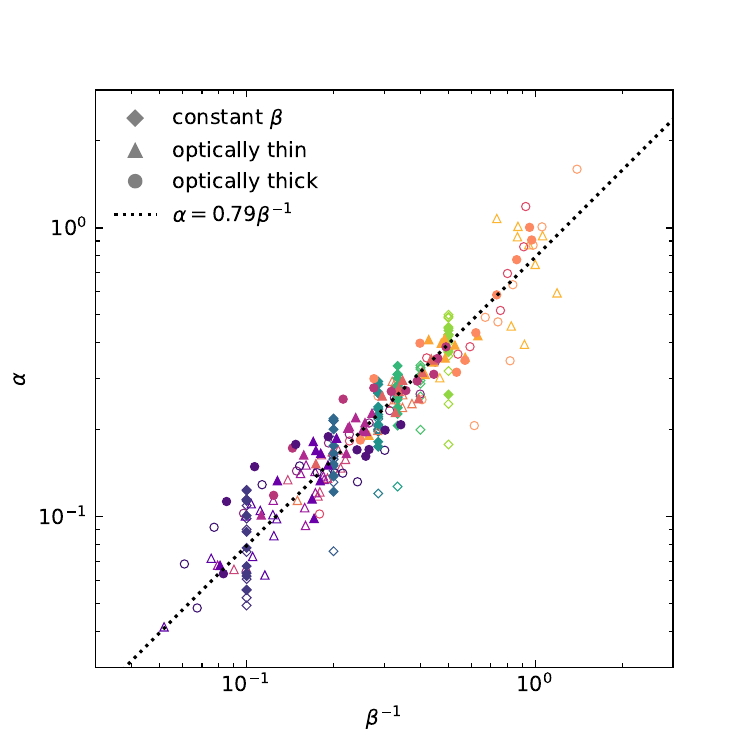}
    \caption{Relation between heating/cooling ($\beta_{\rm heat}$ and $\beta$) and angular-momentum transport ($\alpha$). For each simulation, we divide the range of radii with average $Q_{\rm K}<2$ into 10~intervals, each containing the same amount of time-averaged disk mass. Each data point shows one such interval. Different colors correspond to different simulations, with the same color scheme as in Fig.~\ref{fig:radial_profile}. Empty/full markers correspond to low/fiducial-resolution runs. A linear relation (dotted line) fits the data well.}
    \label{fig:alpha_beta}
\end{figure}

\begin{figure}
    \centering
    \includegraphics[scale=0.66]{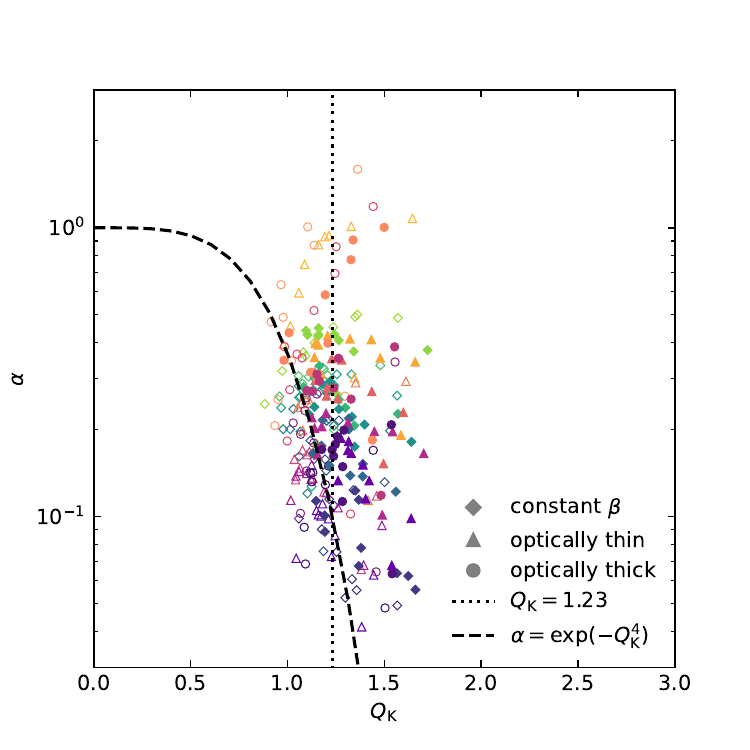}
    \caption{Similar to Fig.~\ref{fig:alpha_beta} but for $\alpha$ vs. $Q_{\rm K}$. There is no clear one-to-one relation between these two quantities, but because all runs lie within a narrow range of $Q_{\rm K}$, a constant $Q_{\rm K}$ (dotted line, with value corresponding to the mean of all points) or any sufficiently steep profile (e.g., the dashed line, based on \citealt{Zhu+10}) can serve as a reasonably good approximations.}
    \label{fig:alpha_Q}
\end{figure}

\begin{figure}
    \centering
    \includegraphics[scale=0.66]{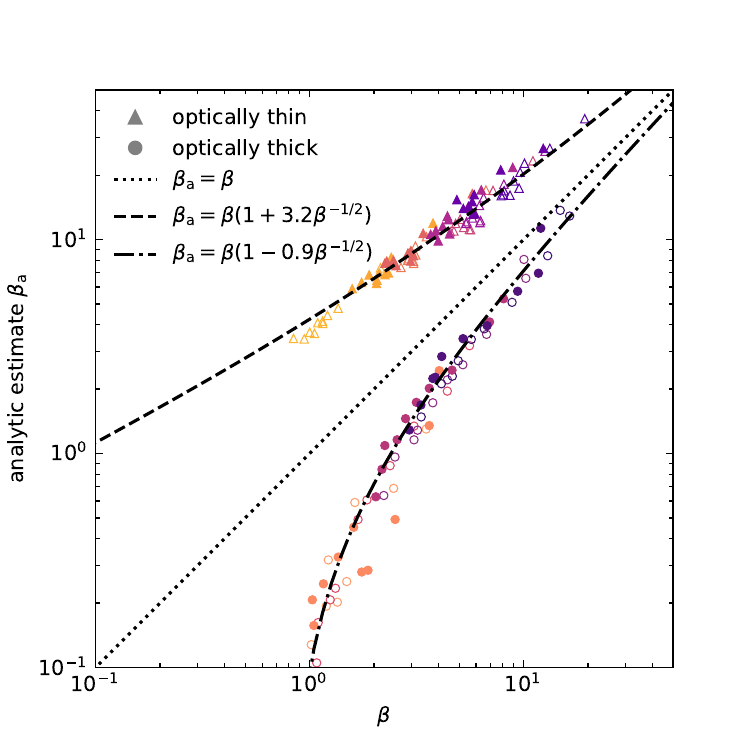}
    \caption{Similar to Fig.~\ref{fig:alpha_beta} for $\beta$ vs. the analytic estimate $\beta_{\rm a}$ based on azimuthally averaged surface density and mean temperature. $\beta$ can show significant deviation from $\beta_{\rm a}$ due to finite perturbation amplitude. To capture the relation between $\beta$ and $\beta_{\rm a}$, we propose a simple formula that includes a correction term proportional to the perturbation amplitude (Eq. \ref{eq:beta_correction}), and it fits the simulation results very well (dashed and dash-dotted lines).}
    \label{fig:beta_analytic}
\end{figure}

\begin{figure}
    \centering
    \includegraphics[scale=0.66]{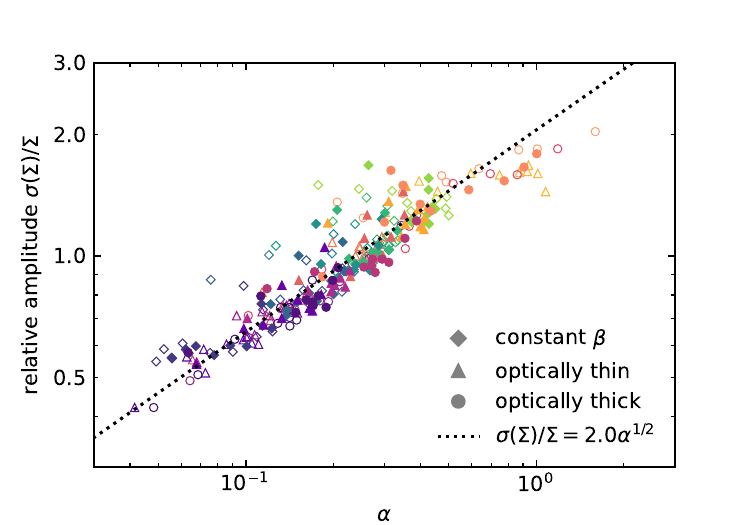}
    \caption{Similar to Fig.~\ref{fig:alpha_beta} for $\alpha$ vs. the amplitude of surface density perturbation, defined with the azimuthal standard deviation $\sigma(\Sigma)$. The perturbation is proportional to $\alpha^{1/2}$ (fit: dotted line), even when it reaches nonlinear amplitudes.}
    \label{fig:alpha_amp}
\end{figure}

Traditionally, in semi-analytic disk models the transport and heating associated with GI is often modeled as a local viscosity whose amplitude depends on the Toomre~$Q$ parameter. Our simulations, which span a wide range of perturbation amplitudes and various cooling regimes, provide a good opportunity to test and improve these prescriptions. Specifically, we can  measure quantitatively how the angular-momentum transport, heating, and cooling in a gravitoturbulent disk deviate from those in a non-turbulent viscous disk model. These measurements allow us to improve the accuracy of semi-analytic models through empirically calibrated correction terms. They also provide more quantitative criteria on when a gravitoturbulence can no longer be accurately captured by a local viscosity.

\subsubsection{Relation between angular-momentum transport and heating}
In the previous subsection, we argued that GI can be approximated as a local viscosity after some time and/or radial average. Here we further quantify the relation between angular-momentum transport and heating. In Fig.~\ref{fig:alpha_beta} we find a relatively tight linear relation of
\begin{equation}
    \alpha \approx 0.8\beta_{\rm heat}^{-1}.\label{eq:alpha_beta}
\end{equation}
For comparison, for a local viscosity we have\footnote{This differs from \citet{Gammie01} Eq.~(20) by a constant factor due to a difference in the definition of $\alpha$; see discussion around Eq.~\eqref{eq:alpha_def}.}
\begin{equation}
    \beta_{\rm heat,visc}^{-1} = \frac 32 (\gamma-1) \alpha = \alpha ~{\rm for}~\gamma=5/3.\label{eq:alpha_beta_visc}
\end{equation}
The difference between Eq.~\eqref{eq:alpha_beta} and Eq.~\eqref{eq:alpha_beta_visc} reflects the $\mathcal O(H/R)$ deviation caused by the anomalous self-gravity energy flux $F^*$.

In the literature, gravitoturbulence is often described as being driven by cooling, with a linear relation between $\alpha$ and $\beta^{-1}$ \citep{Gammie01}. Following this convention, in the bottom panel of Fig.~\ref{fig:alpha_beta} we compare $\alpha$ with $\beta^{-1}$. This results in a linear relation almost identical to the $\alpha$--$\beta_{\rm heat}$ relation, which is not surprising since the disk is near thermal equilibrium with $\beta_{\rm heat}\approx\beta$. We note that the $\alpha$--$\beta_{\rm heat}$ relation is more fundamental than the $\alpha$--$\beta$ relation, since the former is an intrinsic property of the (averaged) gravitoturbulence while the latter holds only under the additional assumptions of thermal equilibrium and GI being the main heating source. 

It is worth noting that the relation between $\alpha$ and $\beta_{\rm heat}^{-1}$ is not sensitive to the cooling type or the amplitude of the turbulence; even fragmenting disks with order-unity $\alpha$ still obey this linear relation. This is consistent with our observation that the radial extent of the perturbations (e.g., Fig.~\ref{fig:periodogram}), which is the main factor controlling the level of deviation from a local viscosity, remain similar across all simulations.

\subsubsection{Relation between $Q$ and the level of turbulence}
In linear theory, the Toomre~$Q$ parameter determines the stability of the disk and the growth rate of the fastest-growing mode \citep{LinShu66}, and it may seem reasonable to assume that it also controls the level of turbulence, which can be characterized by $\alpha$.
However, as we see in Fig.~\ref{fig:alpha_Q}, there is no obvious correlation between $\alpha$ and $Q_{\rm K}$. Across a wide range of $\alpha$, $Q_{\rm K}$ always falls into a relatively narrow range of ${\approx}1$--$1.5$.

One possible interpretation is that the underlying relation between $Q$ and $\alpha$ is so steep that the distribution in Fig.~\ref{fig:alpha_Q} is close to a vertical line plus some random scatter due to finite integration time.
Theoretically, a steep $Q$--$\alpha$ relation appears reasonable. In a saturated gravitoturbulence, the disk on average should lie on the stability boundary, with growth balanced by dissipation through the effective viscosity of the turbulence. (In reality, both growth and dissipation are stochastic, but one may still define a global stability by, for instance, averaging the time derivative of turbulent energy across some temporal and spatial extent.)
This stability boundary would be shifted by the level of turbulence, which acts as an effective viscosity that damps perturbations. But such shift in marginally stable $Q$ is of order
\begin{equation}
    \Delta Q \sim \frac{\omega_{\rm damp}}{\Omega_{\rm K}} \sim \alpha(\lambda/H)^{-2}.
\end{equation}
Here $\omega_{\rm damp}\sim \alpha\Omega_{\rm K}(\lambda/H)^{-2}$ is the damping rate, with $\lambda$ being the spatial scale of perturbation. Previously, we found $\lambda\sim  5H$ (Section \ref{sec:spiral}). The denominator in the first step is $\Omega_{\rm K}$, which is a rough estimate of the derivative of growth rate in $Q$ across the stability boundary. Given this estimate, for all $\alpha\lesssim 1$ the mean $Q$ at saturation should remain approximately constant.

\subsubsection{Deviation in cooling rate due to finite amplitude}\label{sec:transport:cool}
The cooling rate of an unperturbed, internally heated disk can be approximated by the following analytic formula,
\begin{equation}
    \Lambda_{\rm a}(\tau_{\rm mid}, T_{\rm mean}) \equiv \frac{8\sigma_{\rm SB}\tau_{\rm mid}T_{\rm mean}^4}{[1+(0.875\tau_{\rm mid}^2)^{0.45}]^{1/0.45}}.\label{eq:cooling_analytic}
\end{equation}
Here $\Lambda_{\rm a}$ is the cooling per disk area, $\tau_{\rm mid}$ is the midplane optical depth, $T_{\rm mean}$ is the vertically averaged temperature, and $\sigma_{\rm SB}$ is the Stefan-Boltzmann constant.
As discussed in Appendix A of Paper I, this formula is exact in the limits of $\tau\to 0$ and $\tau\to\infty$, and serves as an empirical fit with percent-level precision for intermediate $\tau$.

However, in our simulations the measured cooling rate can deviate significantly from the analytic cooling rate $\beta_{\rm a}=\Omega_{\rm K}U/\Lambda_{\rm a}$ evaluated using the azimuthally averaged disk profile, as shown in Fig.~\ref{fig:beta_analytic}. This deviation is mainly due to the finite amplitude of GI-driven perturbations; due to the nonlinearity of $\Lambda_{\rm a}$,
\begin{equation}
    \overline{\Lambda_{\rm a}(\tau_{\rm mid}, T_{\rm mean})}\neq \Lambda_{\rm a}(\overline{\tau_{\rm mid}},\overline{T_{\rm mean}}).
\end{equation}
To correct for this difference, we expand the cooling rate $\beta^{-1}$ as a function of perturbation amplitude and retain only the lowest-order term, finding that the perturbation amplitude scales as $\delta\Sigma/\Sigma, \delta T_{\rm mean}/T_{\rm mean} \propto\alpha^{1/2}$. This scaling is suggested by both linear perturbation theory and the empirical result from our simulations (Fig.~\ref{fig:alpha_amp}). To first order in perturbation amplitude,
\begin{equation}
    \beta^{-1} \approx \beta_{\rm a}^{-1} (1+f_\alpha \alpha^{1/2}) \approx \beta_{\rm a}^{-1} (1+f_\beta \beta^{-1/2}).
    \label{eq:beta_correction}
\end{equation}
Here $f_\alpha$ is an order-unity factor that depends on how the cooling rate scales with $\Sigma$ and $T_{\rm mean}$, and $f_\beta\approx 0.8^{-1/2}f_\alpha$ following the linear $\alpha$--$\beta^{-1}$ relation in Fig.~\ref{fig:alpha_beta}. Rearranging the terms, we get
\begin{equation}
    \beta_{\rm a} \approx \beta(1+f_\beta \beta^{-1/2}).
\end{equation}
This relation fits the simulations remarkably well (dashed and dash-dotted lines in Fig.~\ref{fig:beta_analytic}).

\subsubsection{Summary}\label{sec:transport:model}
To wrap up this discussion, we use the results from this section to provide recommendations on parameterizing transport, heating, and cooling in semi-analytic disk models.

The angular-momentum transport induced by GI can be approximated, in a time-averaged sense, by a Toomre-$Q$-dependent local viscosity. The $\alpha(Q)$ relation is very steep, so the exact parametrization does not affect the results too much as long as this steepness and the typical $Q$ around which $\alpha$ shows large variation are reproduced. For example, one may use
\begin{equation}
    \alpha_{\rm GI} = \exp(-Q^{-4}),
\end{equation}
or
\begin{equation}
    \alpha_{\rm GI} = \mathcal{H}(Q_0-Q)~{\rm with}~Q_0=1.2,
\end{equation}
where $\mathcal{H}$ is the Heaviside step function.
One can choose whichever format that is more convenient for the application. For example, when considering a local disk patch, the constant-$Q$ closure (with $\alpha$ chosen to match the required accretion rate) is easy to use, whereas when considering time evolution, the smoother exponential profile is often more convenient.

At a given $\alpha_{\rm GI}$, the heating rate produced by GI can be approximated by
\begin{equation}
    \beta_{\rm heat,GI}^{-1} = \frac 32 (\gamma-1) (1+5h)\alpha_{\rm GI}.
\end{equation}
Here $h$ is the aspect ratio $H/R$.
The format of the RHS is motivated by the theoretical argument that the heating rate deviates from the viscous $\alpha$--$\beta_{\rm heat}$ relation by $\mathcal O(h)$, and the factor of 5 is calibrated using Eq.~\eqref{eq:alpha_beta} at $h\approx 0.04$ (Fig.~\ref{fig:radial_profile}).

The finite amplitude of GI-induced perturbations also causes the cooling rate to deviate from that of an unperturbed disk. To first order in perturbation amplitude, we can model the cooling rate as
\begin{equation}
    \beta^{-1} = \beta_{\rm a}^{-1}(1+f_\alpha\alpha_{\rm GI}^{1/2}),
\end{equation}
Here $\beta_{\rm a}$ is the normalized cooling rate of an unperturbed disk, computed from the empirical cooling rate $\Lambda_{\rm a}$ in Eq.~\eqref{eq:cooling_analytic}.
The prefactor of the finite-amplitude correction term $f_\alpha$ can be parametrized as
\begin{equation}
    f_\alpha = 1.2 \frac{\partial\log\beta_{\rm a}}{\partial\log T_{\rm mean}} + 2.9 \frac{\partial\log\beta_{\rm a}}{\partial\log \Sigma}.
\end{equation}
The format of this parametrization is based on the idea that the deviation from $\beta_{\rm a}$ is due to the dependence of $\beta_{\rm a}$ on $\Sigma$ and $T_{\rm mean}$, and the prefactors are calibrated using the optically thin ($\tau_{\rm mid}\to0$) and optically thick ($\tau_{\rm mid}=\tau_0/2=5$; this gives $\partial\log\beta_{\rm a}/\partial\log \Sigma=-1.60$) simulations in Fig.~\ref{fig:beta_analytic} as well as the $\alpha$--$\beta$ relation in Fig.~\ref{fig:alpha_beta}. This parametrization allows a physically motivated extrapolation of our simulation results to other optical depths.
It also allows the temperature dependence of the opacity in a realistic disk to be captured by the ${\partial\log\beta_{\rm a}}/{\partial\log T_{\rm mean}}$ term.

We note that, to the lowest order in $h$ and $\alpha_{\rm GI}$, $\beta_{\rm heat}$ and $\beta$ are reduced to that of a simple viscous disk. However, since the lowest-order corrections have a large prefactor ($5h$) for heating and a small exponent ($\alpha_{\rm GI}^{1/2}$) for cooling, these corrections may cause some nontrivial differences even when $h$ and $\alpha_{\rm GI}$ are well below unity.

\section{Spiral substructures}\label{sec:spiral}
Traditionally, spiral perturbations in a gravitationally unstable disk are often modeled following the formalism of \citet{LinShu66}, in which the perturbation is assumed to be dominated by the fastest-growing linear eigenmode and can therefore be modeled using linear theory. The perturbations then take the shape of grand-design spirals with exact $m$-fold symmetry. This leaves a large but often overlooked gap between theory and simulation: in simulations, the perturbations are seldom, if ever, dominated by a single mode. This is evident in both simulation snapshots, which lack any obvious $m$-fold symmetry (Fig.~\ref{fig:overview_rho}), and in the broad azimuthal spectra showing comparable amplitudes across a wide range of $m$~(Fig.~\ref{fig:periodogram}).
These discrepancies demand an extension or revision to the original Lin \& Shu picture. For example, one could assume that there are multiple unstable modes that each grow independently; or perhaps that the nonlinear evolution of a mode leads to shocks and small-scale turbulence that distort its appearance. These relatively trivial extensions are often implicitly assumed when previous studies use linear theory to explain simulations. But is there any fundamental feature about gravitoturbulence that cannot be captured by a trivial extension of linear theory? The goal of this section is to address this question and provide a more comprehensive physical picture of spiral perturbations in gravitoturbulence that bridges the gap between theory and simulation. This is important not only for conceptualizing gravitoturbulence but also for understanding the observational signatures of GI.
\subsection{Dissecting perturbations: modes are bundled into clumps}
The spiral perturbations show complex dynamics. Qualitatively, the snapshots (Fig.~\ref{fig:overview_rho}) resemble the filamentary perturbations reported in previous studies \citep[e.g.,][]{Nelson+98,Brucy+21}, but it is difficult to form a clear physical picture from the snapshots alone. The periodogram (Fig.~\ref{fig:periodogram}) offers more information on the perturbation in Fourier ($m$) and frequency ($\Omega_{\rm p}$) space, yet it cannot characterize the potential interaction between the modes or the morphology of the perturbations. To combine the strengths of these methods and characterize the morphology of spiral perturbations in more detail, we isolate the perturbations around a given $\Omega_{\rm p}$ and plot them in both real space and Fourier ($m$) space in Fig.~\ref{fig:clump}.
To isolate the perturbations around a given $\Omega_{\rm p}$, we filter the time evolution of surface density perturbation in frequency space using a narrow Gaussian filter centered at this $\Omega_{\rm p}$ with width $0.05\Omega_{\rm p}$.
For each $\Omega_{\rm p}$, the perturbations take the form of one or multiple dense clumps near corotation, with radial and azimuthal extents comparable to the scale height (second and third row). These clumps are often accompanied by spiral perturbations extending further from corotation, but these spirals have significantly lower density compared to the clumps. In Fourier space, each clump consists of a wide range of modes, which are aligned in phase and have similar amplitudes (last two rows).

When there are multiple clumps at the same $\Omega_{\rm p}$ (or the same radius, since clumps are close to corotation), the clumps are close to being evenly spaced (columns~b and~c), but the azimuthal spectrum suggests that they do not show exact $m$-fold symmetry (bottom row). This suggests that these clumps are better interpreted as individual objects.

Comparing the density perturbation of the whole disk (Fig.~\ref{fig:clump}, top-left panel) with individual clumps (second row), we also notice that the disk appears to show a few large-scale spirals/filaments that span the whole disk, but these spirals/filaments do not correspond to individual clumps. This is mainly because multiple clumps can align with each other by chance to create a visually coherent structure. For example, Fig.~14 of \citet{Brucy+21} shows a filament containing multiple density maxima (clumps), each with a width ${\sim}H$. But such alignment is transient, as the clumps have different pattern speeds (see an example in Fig.~10 of \citealt{Bethune+21}). Compared to the large-scale spirals/filaments, clumps are a more atomic (and perhaps more fundamental) unit of gravitoturbulence.

In summary, the perturbations in  gravitoturbulence can be described as a collection of clumpy spirals, each of which consist of a set of aligned, corotating modes with different values of $m$ at similar amplitudes.
The formation of clumps through the alignment of multiple modes is a feature that cannot be explained through linear theory or a trivial extension of it, and hints that the nonlinear coupling between modes plays a crucial role in setting the spiral properties.

\subsection{A theory of clump formation}\label{sec:spiral:theory}
\begin{figure*}
    \centering
    \includegraphics[scale=0.6]{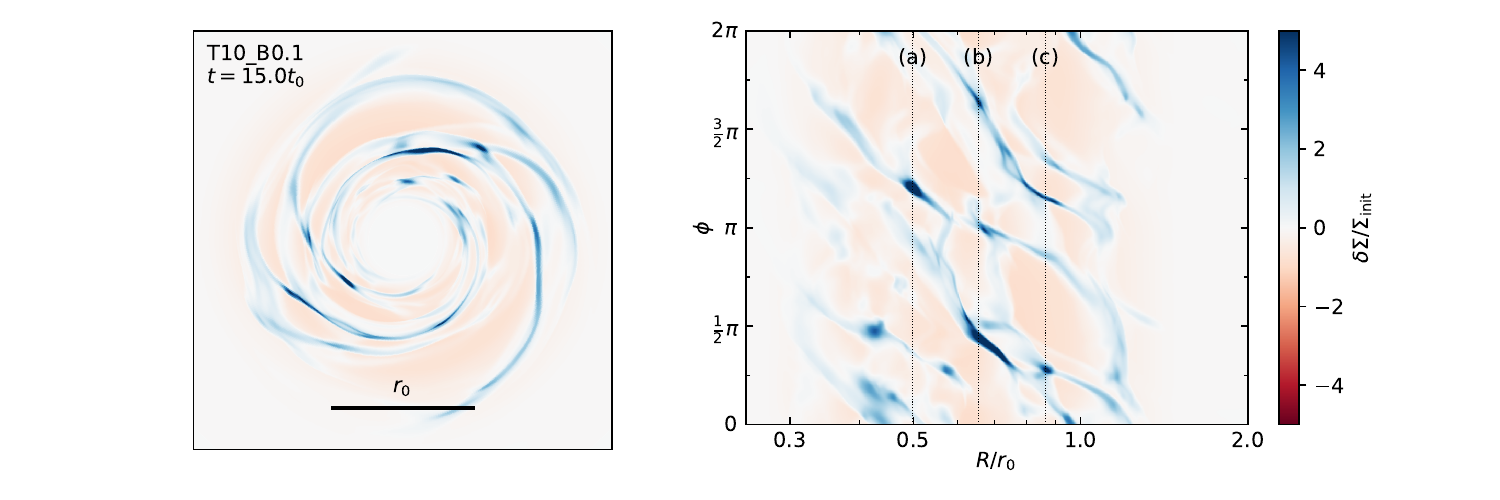}
    \includegraphics[scale=0.6]{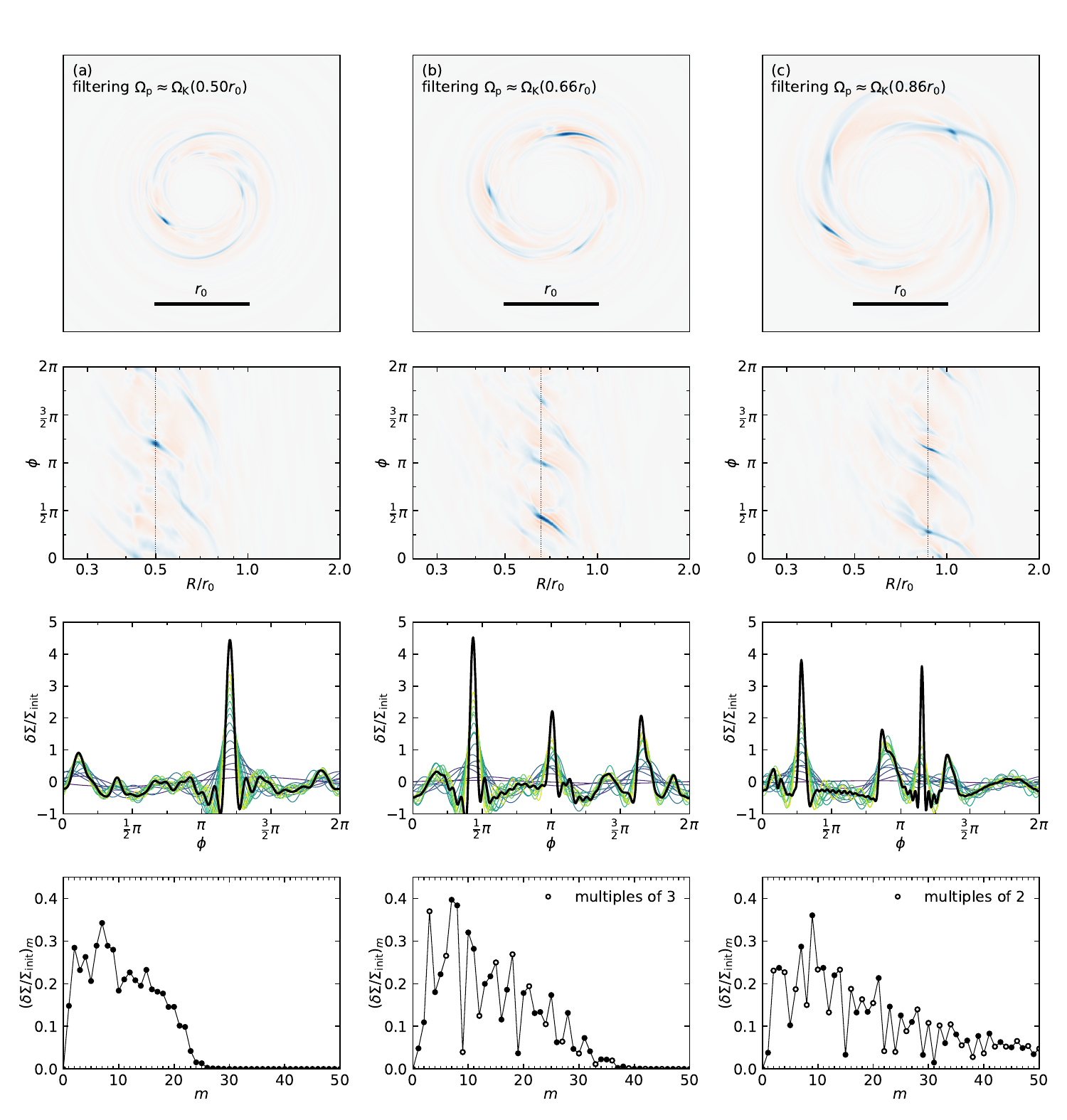}
    \caption{A few clumps isolated from a snapshot, demonstrating the locality of the clumps and how each clump consists of a broad spectrum of aligned, low-amplitude modes. First row: Surface density perturbation $\delta\Sigma/\Sigma_{\rm init}$ from a snapshot in T10\_B0.1. We identify clump locations through local maxima of $\delta\Sigma/\Sigma_{\rm init}$; three of them are marked by vertical lines. In the rest of the figure, each column isolates perturbations corotating with one of these radii by applying a narrow Gaussian filter in $\Omega_{\rm p}$. Each column shows, from top to bottom, the clump in Cartesian and cylindrical coordinate, an azimuthal slice of the clump with colored lines showing contribution from modes with $m\leq m_0$ for $m_0$ between 1 and 20, and the spectrum of this azimuthal slice. For columns (b) and (c), we mark the overtones of $m=3$ and $2$ modes with empty markers to show that they do not dominate the perturbation, suggesting that the multiple clumps at a given radius are better interpreted as individual objects rather than a single object with $m=3$ or $2$ symmetry.}
    \label{fig:clump}
\end{figure*}

\begin{figure}
    \centering
    \includegraphics[scale=0.66]{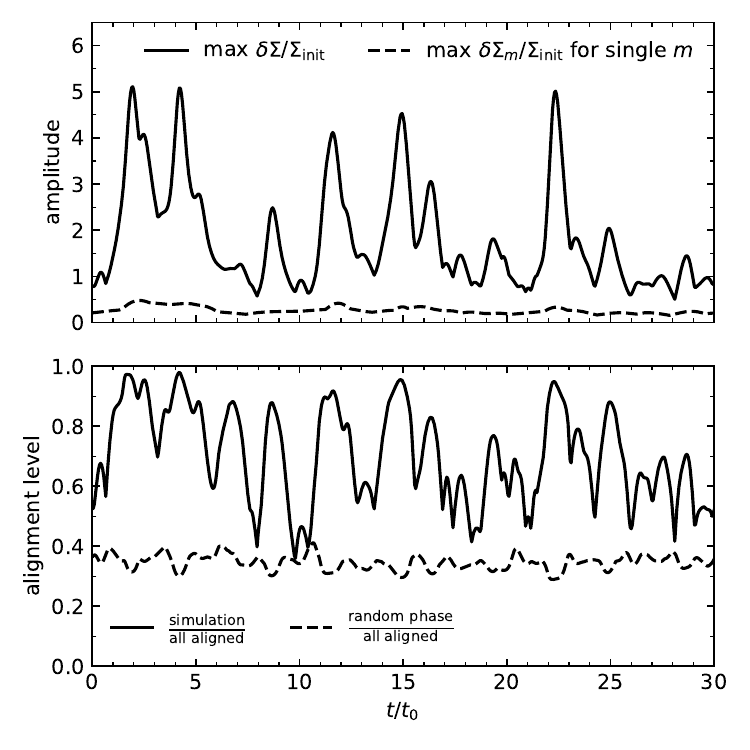}
    \caption{Time evolution of perturbation amplitude (top panel) and alignment level (bottom panel) at $r=0.5r_0$ for perturbations with $\Omega_{\rm p}\approx \Omega_{\rm K}(0.5r_0)$ in run T10\_B0.1. (This is the perturbation component in column (a) of Fig.~\ref{fig:clump}.) In the bottom panel, we compare the maximum $\delta\Sigma/\Sigma_{\rm init}$ with the maximum $\delta\Sigma/\Sigma_{\rm init}$ when all Fourier ($m$) components are aligned (solid line) and the median value of the maximum $\delta\Sigma/\Sigma_{\rm init}$ when all Fourier components have random phases (dashed line) to characterize the level of alignment. The nonlinear coupling between modes increases alignment, especially when the perturbation amplitude is large.}
    \label{fig:mode_alignment}
\end{figure}

The finding in the previous subsection prompts many questions. Why do we see excitation of modes across a wide range of $m$, and why are they all aligned? What sets the typical size and amplitude of a clump? Here, we provide some physical arguments to answer these questions.

In Fourier space, a clump is a wide spectrum of aligned modes (Fig.~\ref{fig:clump}); two factors contribute to the formation of these modes. First, all tightly wound modes ($m\lesssim R/H$) share approximately the same growth rate; the relative deviation from a constant growth rate is only $\mathcal O[(mH/R)^2]$ \citep{LinShu66,LauBertin78}. Second, the nonlinear coupling between modes must be non-negligible in a saturated gravitoturbulence because it is required to saturate the linear growth of modes \citep{LaughlinRozyczka96} and balance cooling by dissipating energy via turbulent cascades or shocks.
Nonlinear effects both cap the amplitude of individual modes and seed the growth of other unstable modes, causing all modes with $m\lesssim R/H$ to reach similar magnitude.
Meanwhile, since the nonlinear coupling between modes preferentially excites new perturbations that are in phase with existing perturbations, it causes mode alignment.
Together, this picture explains the clumps that are local in $\phi$ with width $\mathcal O(H)$.

We can also follow the above picture to provide an analytic estimate of the amplitude of the modes and the clumps. The typical mode amplitude corresponds to when nonlinear coupling saturates mode growth.
For a mode $m$, the leading-order nonlinear effect consists of coupling with two other modes $m_1,m_2$ satisfying $m=m_1\pm m_2$. Among the modes with nontrivial amplitudes ($m\lesssim R/H$), there are ${\sim} R/H$ pairs of such modes. Therefore, a mode amplitude of $\mathcal O(H/R)$ is sufficient for nonlinearly saturating mode growth. Clumps, which consist of ${\sim}R/H$ aligned modes, then have $\mathcal O(1)$ amplitude.
This estimate is broadly consistent with the amplitudes we find in our simulations (e.g., Fig.~\ref{fig:clump}). One caveat is that our simulations all have similar $H/R$, so we cannot directly test if our proposed $H/R$ scaling of mode and clump amplitude is correct. This needs to be tested in future simulations.

Mode saturation and alignment via nonlinear coupling are visible in our simulations. In Fig.~\ref{fig:mode_alignment} we plot the time evolution of the perturbation amplitude and the level of mode alignment at a given radius for one of our simulations. Mode saturation is visible from the maximum amplitude of individual modes, which remains at 0.2--0.5 (dashed line in the top panel; also see Fig.~\ref{fig:clump}). This is much lower than the total amplitude, suggesting that multiple modes coexist at all times. To demonstrate the alignment between modes, we show that the maximum density perturbation ($\delta\Sigma/\Sigma_{\rm init}$) is similar to the maximum $\delta\Sigma/\Sigma_{\rm init}$ if all modes were perfectly aligned, and is higher than the typical maximum $\delta\Sigma/\Sigma_{\rm init}$ when modes have random phases (bottom panel). We note that alignment persists even when there is no dense clump present (i.e., low maximum $\delta\Sigma/\Sigma_{\rm init}$).

Finally, the radial locality of clumps and the individual modes that comprise the clump can be explained using dissipation via shocks. Clumps have $\mathcal O(1)$ amplitude and a pattern speed close to corotation, so the velocity perturbation associated with them becomes transonic at a distance of ${\sim}H$. Dissipation via shocks thus reduces the amplitude beyond a width of ${\sim}H$ around corotation, creating the localized mode amplitude profiles we see in Fig.~\ref{fig:periodogram}. Additionally, since linear modes with different $m$ have different pitch angles, the collection of modes with $m=1\sim R/H$ that are aligned at corotation become out of phase with each other a radial distance ${\sim}H$ away from corotation. This makes the clump even more localized in real space.

\subsection{When does GI produce grand-design spirals?}\label{sec:spiral:grand-design}

So far we have focused on the regime where GI produces clumpy spirals, which is seen in all our simulations. But does there exist a regime where GI produces grand-design spirals with $m$-fold symmetry?

Previous studies have argued that the spiral features become more global at larger disk masses. This feature can be attributed to the larger $H/R$ in these disks: even if the spirals are qualitatively the same as those in the thinner disks, the ${\sim} H$ clumps are now so large that they appear like global structures. On the other hand, there is no convincing evidence for whether the spirals make a qualitative switch to symmetric grand-design spirals at sufficiently large disk mass (or $H/R$). Theoretically, it might make sense that eventually so few modes have large enough linear growth rates (which requires $m\lesssim R/H$) that the perturbations become dominated by a single mode. But it is unclear whether such a switch occurs in simulations. For example, in most snapshots in \citet{LodatoRice05} and \citet{Forgan+11}, the spirals lack apparent $m$-fold symmetry even for a disk-to-star mass ratio of $0.5-1.5$. Meanwhile, in an attempt to reproduce the symmetric grand-design spirals observed in Elias~2-27 by \citet{Hall+18}, some simulated spirals do show $m=2$ symmetry, but there is no obvious correlation between disk mass and spiral morphology.

The physical picture we developed regarding clump formation offers an alternative hypothesis that can be tested with future simulations.
In Section~\ref{sec:spiral:theory} we argue that perturbations always become local clumps in a saturated gravitoturbulence, as saturation mandates a nontrivial level of nonlinear mode-mode coupling. In other words, global spirals occur only when GI is not in a saturated state. For example, this may occur when the disk suddenly becomes more unstable due to episodic envelope infall, or during the stage when the disk has become gravitationally stable and the gravitoturbulence is decaying with different modes decaying at different rates. This may explain why global spirals that quantitatively agree with a single linear eigenmode of GI are sometimes seen in observation \citep[e.g.,][]{Paneque-Carreno+21,Xu+23} but usually not in simulations of saturated gravitoturbulence \citep[e.g.,][and this work]{Cossins+09,Bethune+21,Steiman-Cameron+23}.
Meanwhile, symmetric grand-design spirals can be produced in simulations when GI is not in a saturated state; for example, \citet{LodatoRice05} report that a $m=2$ grand-design spiral appears during the initial transient before reaching a saturated gravitoturbulence. In \citet{Rowther+24}, when GI is suppressed by stellar heating, the decaying gravitoturbulence is not only weaker but also more symmetric compared to an active, saturated gravitoturbulence.

\section{Observational signatures of spiral substructures}\label{sec:spiral:obs}
Gravitoturbulence (at least as realized in our simulations) consists of modes occurring at a wide range of $m$ that bundle into clumpy spirals. How does this bundling affect the observational properties of the spirals? To what extent do the insights derived from models assuming symmetric grand-design spirals consisting of a single linear eigenmode still hold?
In this subsection, we provide some intuition on these issues using our simulation data. One caveat is that we only consider optically thin tracers of column density and usually ignore instrumental effects. In reality, the spiral substructures could have different appearences depending on the abundance and optical depth of the tracer as well as the limited sensitivity and resolution of the instrument \citep[e.g.,][]{Dipierro+14,Dipierro+15,Dong+15,Hall+16,Hall+18,Hall+19}. The investigation of how these factors affect our results is left for future studies.

\subsection{Qualitative spiral morphology}

\begin{figure*}
    \centering
    \includegraphics[width=\textwidth]{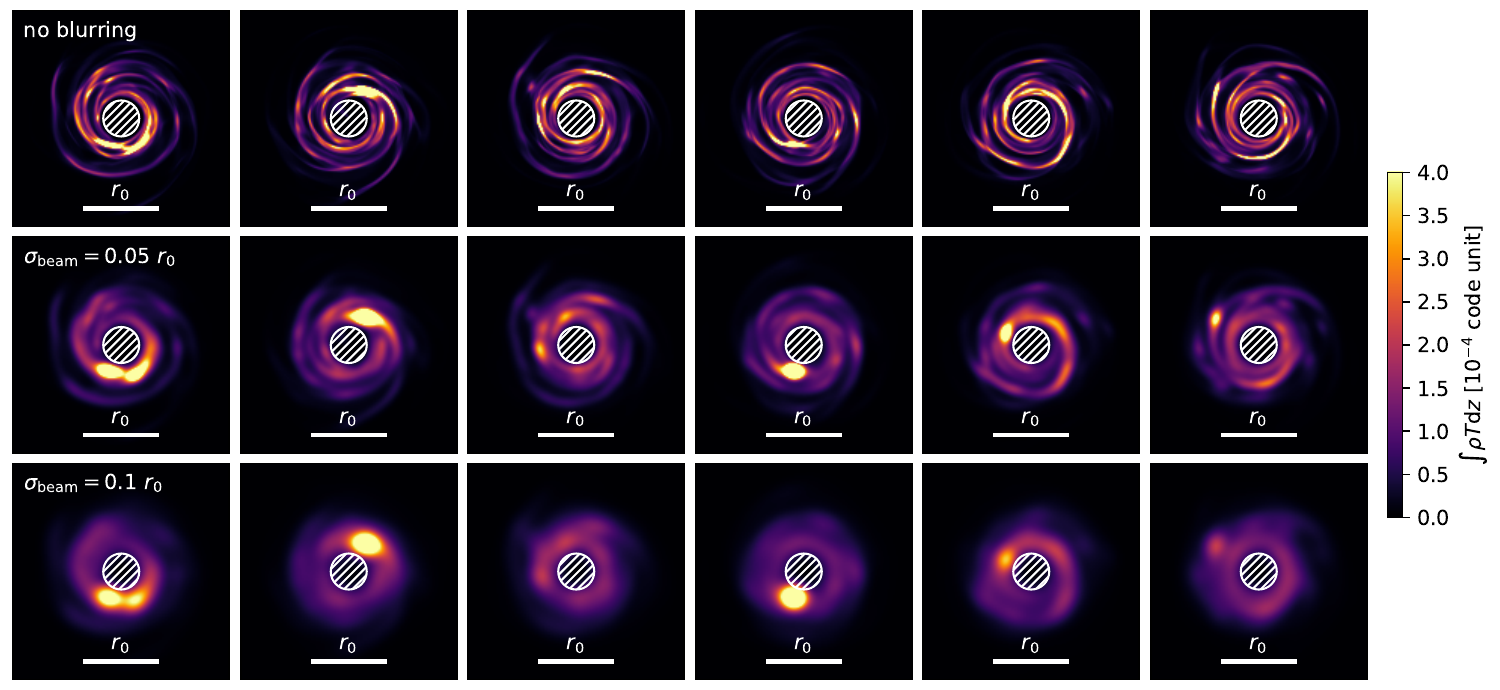}
    \caption{Vertically integrated $\rho T$ from simulation T10\_B0.3. This is directly proportional to the dust continuum intensity if at the wavelength of observation the emission is optically thin and in the Rayleigh-Jeans limit. We take six different epochs located uniformly between 15 and 30 $t_0$, and for epoch we apply different levels of Gaussian blurring to mimic the effect of finite resolution. Although the simulation contains many spirals, we see few or no spirals once the beam size is $\gtrsim H$. Instead, the emission is often dominated by a single clump, and that may cause the substructure to be misidentified as a binary companion or embedded planet. Note that no fragmentation occurs in this simulation and all clumps are transient.}
    \label{fig:spiral_obs}
\end{figure*}

Global spirals that correspond to single linear eigenmodes form ``grand-design'' spirals with exact $m$-fold symmetry; this is also often assumed to be a signature of GI \citep{Forgan+18}.
For the clumpy spirals seen in simulations,
although some spirals may still appear global due to the accidental alignment of multiple clumps, the spirals generally show no $m$-fold symmetry.

The observational properties of clumpy spirals can also be sensitive to resolution.
In Fig.~\ref{fig:spiral_obs} we take one of our simulations and plot the vertically integrated $\rho T$ at a few different epochs. This is proportional to the dust continuum intensity if at the observed wavelength the emission is optically thin and in the Rayleigh-Jeans limit.
We find that, under finite resolution (middle and bottom rows), the observed perturbation is often dominated by bright clumps instead of spirals \citep[also see similar results in][]{Dipierro+14}. Several factors contribute to this. First, most of the perturbation energy and amplitude is concentrated in a few scale heights around corotation (Fig.~\ref{fig:periodogram}), causing the spirals around the clumps to be much dimmer than the clumps (Fig.~\ref{fig:clump}). Second, since the clumps/spirals show little coherence, when multiple spiral substructures are combined by finite resolution, they often become irregular asymmetries. In contrast, for a single linear eigenmode (single $m$), low resolution decreases the amplitude but does not change the morphology (e.g., Fig.~4 in \citealt{Xu+23}).

In summary, these features pose three challenges in identifying the local clumps produced by GI: resolution, sensitivity, and misclassification. The low amplitude of spirals and their lack of coherence pose more stringent requirements in resolution and sensitivity compared to global spirals that correspond to a single linear eigenmode.\footnote{The challenge in sensitivity is further amplified by the fact that the outer disk, where perturbations show higher amplitude and are easier to resolve, tend to by much dimmer than the inner disk; see discussion in \citet{Xu+23}.} Meanwhile, the local clumps that dominate the perturbation can easily be misidentified as noise (e.g., from interferometry image reconstruction) or embedded planet/binary companions \citep[e.g,][]{Currie+22,Tobin+22}.
This last issue calls for more systematic studies in distinguishing between GI-produced (often non-fragmenting) clumps and actual planet/binary companions.

\subsection{Pitch angle}\label{sec:obs:pitch}

\begin{figure}
    \centering
    \includegraphics[scale=0.66]{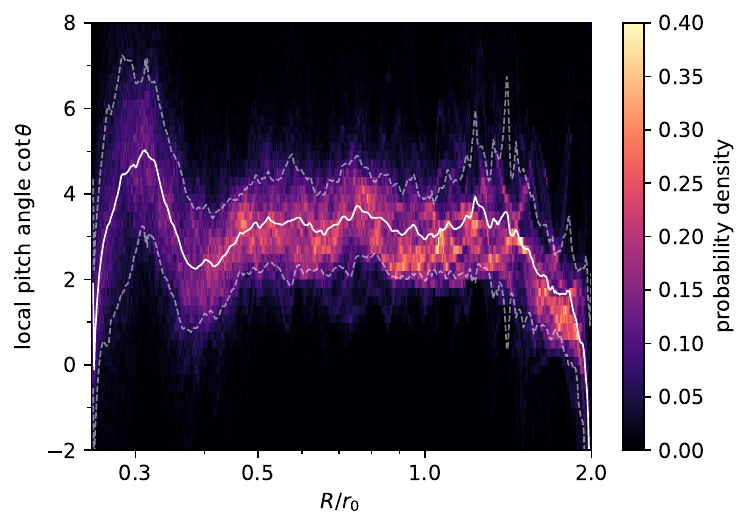}
    \caption{The distribution of local pitch angle at each radius for run run T10\_B0.1. Mean and $\pm1\sigma$ values are shown in solid and dashed lines for reference. In the gravitationally unstable region ($R/r_0\approx0.4-1$; see Fig. \ref{fig:radial_profile}) the pitch-angle distribution is approximately the same, with $\cot\theta\approx 3\pm1$.}
    \label{fig:pitch}
\end{figure}

\begin{figure}
    \centering
    \includegraphics[scale=0.66]{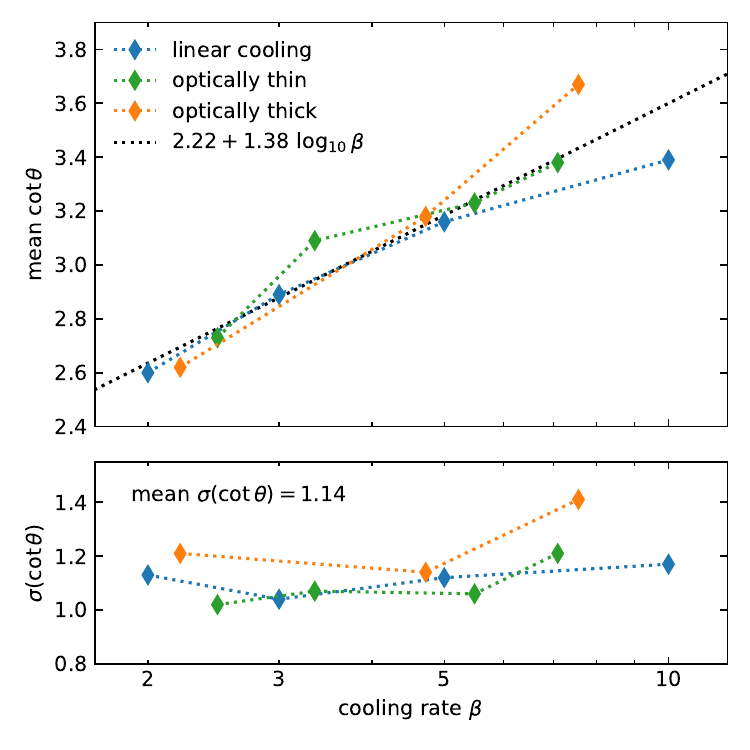}
    \caption{The mean and standard deviation of the pitch-angle distribution for our simulations. The horizontal axis is the mean cooling timescale. Mean $\cot\theta$ shows a dependence on cooling rate, with slower cooling giving more tightly wound spirals; a log-linear fit is given in the top panel (black dotted line). The width of the $\cot\theta$ distribution remains similar across simulations. For all simulations, we only use the gravitationally unstable region $R/r_0\in[0.4,1]$ to evaluate the distribution of $\cot\theta$.}
    \label{fig:pitch_cooling_dependence}
\end{figure}

\begin{figure}
    \centering
    \includegraphics[scale=0.66]{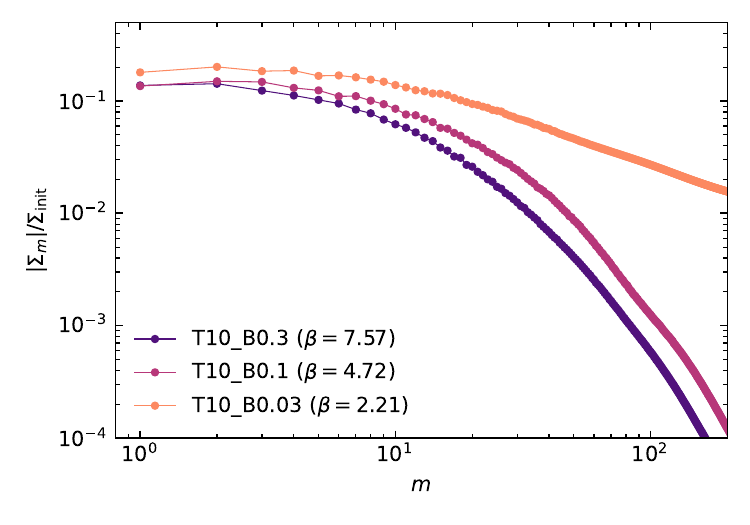}
    \caption{Azimuthal spectrum of density perturbation for optically thick runs. The broad spectrum is primarily due to the presence of clumps with size $\sim H$. Faster cooling increases the level of clump contraction, resulting in broader spectra. Other cooling types show the same qualitative trend.}
    \label{fig:spectrum}
\end{figure}

Pitch angle is an important observational property of spiral perturbations. For grand-design spirals consisting of a single linear eigenmode, the pitch angle can be estimated using the linear dispersion relation. The fastest-growing linear mode has a radial wavenumber
\begin{equation}
    k_R \approx \frac{\pi G\Sigma}{c_{\rm s,2D}^2+2\pi G\Sigma H}.\label{eq:k_R}
\end{equation}
Here $c_{\rm s,2D}$ is the effective sound speed for 2D perturbations. We discuss the origin of this formula in Appendix~\ref{a:pitch}. For a given $Q$, $k_R \propto 1/H$. The resulting pitch angle is
\begin{equation}
    \cot\theta = \frac{k_R R}{m} \propto \frac{R}{mH}.
\end{equation}
The pitch angle depends on $m$ and $H/R$, but not on the cooling rate.

Meanwhile, the typical pitch angle of local clumps show a different trend.
In Fig.~\ref{fig:pitch} we evaluate the local pitch angle of perturbations in our simulation and plot the resulting distribution. The local pitch angle at a given time and radius is defined as follows. Consider the surface density profile $\Sigma(r,\phi)$ at two adjacent radial cells, $r$ and $r+\Delta r$. The azimuthal phase shift between the cells $\Delta\phi$ can be determined by maximizing $\int \Sigma(r,\phi+\Delta\phi)\Sigma(r+\Delta r,\phi)\,{\rm d}\phi$. The pitch angle $\theta$ is then given by
\begin{equation}
    \cot\theta = \Delta\phi/\Delta \log r.
\end{equation}
Where $\Delta\log r = \log(1+\Delta r/r)$.
Within the gravitationally unstable disk, we find a typical pitch angle $\cot\theta \approx 3$ with a $1\sigma$ spread of ${\approx} 1$.
Fig.~\ref{fig:pitch_cooling_dependence} compares the pitch-angle distribution across different simulations.
The typical $\cot\theta$ slightly increases (i.e., more tightly wound) with increasing $\beta$ (slower cooling). Meanwhile, at a given $\beta$ the pitch angle is not sensitive to the cooling type.

One caveat is that the pitch angle also shows some resolution dependence, with low-resolution simulations showing more tightly wound spirals. We discuss the origin of this resolution dependence in Appendix~\ref{a:res} and use additional tests to demonstrate that the pitch angle has likely reached numerical convergence around our fiducial resolution.

The cooling rate dependence of the pitch angle can be interpreted as a result of varying levels of clump contraction. When cooling is dynamically unimportant ($\beta\ll 1$), pressure support stops the gravitational contraction of clumps at ${\sim}H$, as we discussed in the previous subsection. But for faster cooling, the clump can contract further; eventually, for sufficiently fast cooling, fragmentation can occur, as we discussed in Paper~I. The higher level of contraction means that modes with higher $m$ can be excited. This is visible when we compare the spectrum of runs at different cooling rates (Fig.~\ref{fig:spectrum}). The spectra of individual clumps in Fig.~\ref{fig:clump} show the same trend: for this simulation, $\beta$ decreases outwards (see Fig.~\ref{fig:radial_profile}), and clumps at larger radii show excitation of higher $m$ modes.
Meanwhile, for each mode, the radial wavenumber $k_R$ remains approximately constant (Eq.~\ref{eq:k_R}; also see \citealt{Cossins+09} and Appendix~\ref{a:pitch}), which corresponds to $\cot\theta = k_R R/m$ decreasing with $m$. As a result, lower $\beta$ and further clump contraction means the inclusion of more loosely wound modes, reducing the typical $\cot\theta$ as we see in Fig.~\ref{fig:pitch_cooling_dependence}.
Under this picture, we also expect that the pitch angle will be insensitive to $H/R$. This is because the clump size scales with $H$ and the typical pitch angle of modes in this clump corresponds to $m\sim R/H$ and $k_R\sim$ constant.

In summary, for grand-design spirals (single mode) the pitch angle scales with $H/R$ and $m$, whereas for clumpy spirals (multiple aligned modes) the pitch angle is a distribution of finite width with the typical pitch angle set by the cooling rate.

\subsection{Kinematic signature}

\begin{figure}
    \centering
    \includegraphics[scale=0.66]{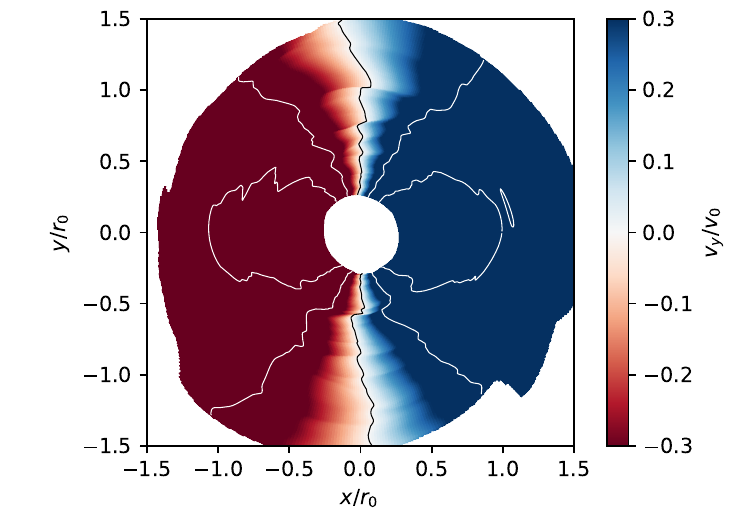}
    \caption{Density-weighted vertical average of $v_y/v_0$ from a snapshot of simulation T10\_B0.3. Here $v_0$ is the Keplerian speed at $r_0$. This average $v_y$ serves as a rough approximation of the line-of-sight velocity (up to a $\sin I$ factor) for a near face-on disk with $y$ being the minor axis. We only plot the region with $\Sigma>10^{-2}\Sigma_{\rm init}$. Lines show isocountours at $0,\pm0.5$ and $\pm 1v_0$. The GI wiggle, which is the perturbation to the isocountours of line-of-sight velocity due to GI-driven perturbations, is clearly visible.}
    \label{fig:GI_wiggle}
\end{figure}

GI also produces a kinematic signature known as ``GI wiggle'' \citep{Hall+20}: the transonic velocity perturbation of GI produces wiggles in the isocountours of the line-of-sight velocity, which translates to wiggles in channel maps. Fig.~\ref{fig:GI_wiggle} demonstrates similar wiggles in our simulation.

GI wiggle can be modeled quantitatively to constrain disk properties and the level of turbulence \citep{Longarini+21,Terry+22}. However, such modeling assumes that perturbations come from a single linear eigenmode; does such a model still produce useful constraints when perturbations take the form of local clumps, as in our simulations? In both cases, the length scale of the GI wiggle constrains the scale height and the amplitude constrains the turbulent velocity, which then constrains $\alpha$. As a result, such modeling remains correct up to some order-unity error. Meanwhile, it is worth pointing out that such modeling only constrains the instantaneous $\alpha$, and translating such instantaneous $\alpha$ to cooling rate or accretion rate will inevitably produce a large uncertainty due to the strong variability of $\alpha$ (Section~\ref{sec:transport:visc}). In summary, modeling GI wiggles as a single linear eigenmode can still provide a rough estimate of scale height, turbulence strength, and cooling rate, but the uncertainty is at least of order unity.

\section{Conclusion}\label{sec:conclusion}
In this paper, we use 3D global hydrodynamic and radiation hydrodynamic simulations to investigate the spiral perturbations in gravitationally unstable accretion disks. Below we summarize our main findings.

\begin{itemize}

\item Most spiral properties are insensitive to cooling type (constant $\beta$ or radiative, optically thin or optically thick) and are mainly determined by the cooling rate (effective $\beta$). The main exception is that radiative cooling, especially in the optically thick regime, reduces the amplitude of temperature perturbations (Section~\ref{sec:overview}).

\item The instantaneous transport of GI cannot be modeled as a local viscosity, but the time-averaged transport of GI is equivalent to a local viscosity to zeroth order in $H/R$ and $\alpha^{1/2}$ (Section \ref{sec:transport}). We develop prescriptions for the transport, heating, and cooling in a gravitoturbulent disk up to first order in $H/R$ and $\alpha^{1/2}$ in Section~\ref{sec:transport:model}. These prescriptions can be applied to semi-analytic modeling of GI in protostellar disks.

\item The spiral perturbations in  gravitoturbulence are clumpy (Section~\ref{sec:spiral}). The perturbations consist of a wide range of modes up to $m\sim R/H$, and modes at a given pattern speed tend to be aligned. In real space, these aligned modes form dense clumps (with size ${\sim}H$ in both the radial and azimuthal directions) accompanied by low-amplitude spirals.
Clumping through the alignment of modes is a feature not inherent in the traditional Lin \& Shu picture of spiral perturbations, and likely originates from the nonlinear coupling between modes.

\item Clumping is a generic feature of gravitoturbulence, and is not exclusive to fragmenting disks. Together with our results from Paper I, this suggests a continuous transition between gravitoturbulence and fragmentation: clumping occurs universally (Section \ref{sec:spiral:theory}), but clumps evolve into bound fragments before being destroyed by clump--clump collisions only when cooling is sufficiently strong (Paper I Section 4).

\item The clumpy spirals seen in simulations deviate significantly from the traditionally assumed symmetric grand-design spirals, which consist of a single linear mode (Section \ref{sec:spiral:obs}). Morphologically, clumpy spirals do not exhibit $m$-fold symmetry, and the clumps are easily misidentified as observational noise or embedded companions. Meanwhile, the kinematic signature of ``GI wiggles'' remains qualitatively (though not quantitatively) similar in both regimes, making it a good option for detecting the presence of GI. The existence of symmetric grand-design spirals might require conditions other than a saturated gravitoturbulence, such as recently triggered or decaying GI (Section \ref{sec:spiral:grand-design}).

\end{itemize}

\section*{}
W.X. thanks Philip Armitage, Hongping Deng, Ruobing Dong, Giuseppe Lodato, and Farzana Meru for insightful discussions. We thank the anonymous referee for providing constructive comments.
Simulations in this work are performed with computational resources at the Flatiron Institute.

\appendix
\twocolumngrid

\section{Radial wavenumber from linear dispersion relation}\label{a:pitch}

The detailed properties of linear eigenmodes in a gravitationally unstable disk need to be computed numerically using global models. However, some insights can still be obtained from the local dispersion relation using the WKBJ approximation, which is given by \citep[see Eqs.~15 and 16 in][]{KratterLodato16}
\begin{equation}
    (\omega- m \Omega)^2 = (c_{\rm s,2D}^2 + 2\pi G\Sigma H) k_R^2 - 2\pi G \Sigma |k_R| + \kappa^2.\label{eq:spiral_dispersion}
\end{equation}
Here $\omega$ is the (complex) frequency of the mode, $\Omega$ and $\kappa$ are the rotation and epicyclic frequencies, and $c_{\rm s,2D}$ is the 2D effective sound speed given by \citep[cf.][]{Goldreich+86}
\begin{equation}
    c_{\rm s,2D}^2 = c_{\rm s}^2\frac{3\gamma-1}{\gamma(\gamma+1)},
\end{equation}
with $\gamma$ being the adiabatic index of the gas.

Eq.~\eqref{eq:spiral_dispersion} is the standard WKBJ dispersion relation for a 2D (zero-thickness) self-gravitating disk \citep{LinShu66}, plus a correction term $2\pi G\Sigma H k_R^2$ that accounts for the finite thickness of the disk \citep{Vandervoort70}.
Note that this relation assumes tight winding, or $k_R R/m\gtrsim 1$; this is usually a reasonable approximation for the simulated and observed GI spirals in protostellar/protoplanetary disks.

Now consider the most unstable mode in Eq.~\eqref{eq:spiral_dispersion}. That requires minimizing the RHS of the equation, which yields
\begin{equation}
    |k_R| = \frac{\pi G\Sigma}{c_{\rm s,2D}^2 + 2\pi G\Sigma H}.\label{eq:kr_analytic}
\end{equation}
This gives an estimate for the pitch angle of the spiral. For a $Q\sim 1$ disk, the two terms in the denominator have similar magnitude, so the finite thickness correction plays a nontrivial role.

The argument above contains two important errors. First, for an unstable mode, both sides of the equation are negative real numbers, requiring that the real part of the frequency corresponds to corotation, ${\rm Re}(\omega)=m\Omega$. That is, however, a condition that cannot be everywhere satisfied within a mode. A more fundamental error is that this linear dispersion relation is too stable: it never yields instability for $Q>1$. In contrast, in both simulations (such as this work) and global linear analyses \citep[e.g.,][]{Chen+21}, unstable spirals exist at $Q>1$.

Still, somewhat surprisingly, this pitch angle argument holds up very well against more accurate linear calculation and observation. In \citet{Chen+21}, the pitch angle derived following this WKBJ argument deviates from the result of global linear analysis by less than 20\% for a power-law disk profile, and correctly predicts how the pitch angle scales with disk properties. (They only consider 2D disks, so their comparison does not contain the finite thickness correction term.) In \citet{Xu+23}, Eq.~\eqref{eq:kr_analytic}, together with the radial temperature and surface density profiles inferred from dust continuum intensity, is used to predict the pitch angle of an observed spiral, and the prediction agrees well with the observed morphology.

Why Eq.~\eqref{eq:kr_analytic} seems to serve as a good estimate calls for future study, and here we offer two conjectures. One possibility is that, while the WKBJ linear dispersion relation does not produce the correct stability result, it still manages to capture the relative level of stability between modes. In other words, minimizing the RHS of Eq.~\eqref{eq:spiral_dispersion} still makes sense. This argument would hold if the additional effect of the global spiral profile is equivalent to adding a term to Eq.~\eqref{eq:spiral_dispersion} that is insensitive to $k_R$. For example, following the insight of \citet{Toomre81}, swing amplification may be the main mechanism of global instability in a WKBJ-stable disk, and its effect is similar to reducing the epicyclic frequency.

Another possibility is that Eq.~\eqref{eq:kr_analytic} should be considered not as a result of minimizing the RHS of the WKBJ dispersion relation, but as a constraint from the imaginary part of the dispersion relation. Physically, this means considering the requirement that the perturbation produced by an eigenmode needs to be in phase with the mode itself. To do this, we generalize the dispersion relation to allow a complex $k_R$, where the imaginary part captures the radial variation of the amplitude. Generalizing into complex $k_R$ is trivial, except for the $|k_R|$ term. To properly handle this term, consider the derivation of the linear dispersion relation following \citet{BinneyTremaine08}, Section 6.2.2. The $|k_R|$ in Eq.~\eqref{eq:spiral_dispersion} is in fact $k_R^2/k_z$, where $k_z$ comes from the gravitational perturbation of form $\delta\Phi\propto \exp(ik_RR-k_z|z|)$. For real $k_R$, the Possion equation above the disk demands $k_z = |k_R|$ (see their Eqs.~5.158--5.161; our $k_R$ corresponds to their $k_x$). For complex $k_R$, the result is similar, except now we have $k_z = {\rm sgn}[{\rm Re}(k_R)]k_R$. As a result, $|k_R|$ in Eq.~\ref{eq:spiral_dispersion} should be replaced by
\begin{equation}
    \frac{k_R^2}{k_z} = {\rm sgn}[{\rm Re}(k_R)]k_R.
\end{equation}
Taking the imaginary part of the dispersion relation, we get
\begin{multline}
    2{\rm Im}(\omega)[{\rm Re}(\omega)-m\Omega] = 2(c_{\rm s,2D}^2 + 2\pi G\Sigma H){\rm Re}(k_R){\rm Im}(k_R) \\- 2\pi G\Sigma~{\rm sgn}[{\rm Re}(k_R)]{\rm Im}(k_R).
\end{multline}
The fastest-growing global mode typically shows a significant amount of radial variability, with $|{\rm Im}(k_r)|\sim H^{-1}$ (e.g., Fig.~2 in \citealt{Chen+21}). Physically, this is because they may be considered as the superposition of leading and trailing waves that feed each other via reflections and overreflections (including swing amplification) at boundaries and resonances \citep{Toomre81}, and the beating between them produces the radial variation of wave amplitude. Meanwhile, the growth rate ${\rm Im}(\omega)$ should usually be small, because the feedback from GI via heating and angular-momentum transport drives the disk towards marginal instability. This allows us to ignore the LHS and obtain
\begin{equation}
    |{\rm Re}(k_R)| \approx \frac{\pi G\Sigma}{c_{\rm s,2D}^2 + 2\pi G\Sigma H}.
\end{equation}

\section{Resolution dependence of pitch angle}\label{a:res}

\begin{figure}
    \centering
    \includegraphics[scale=0.66]{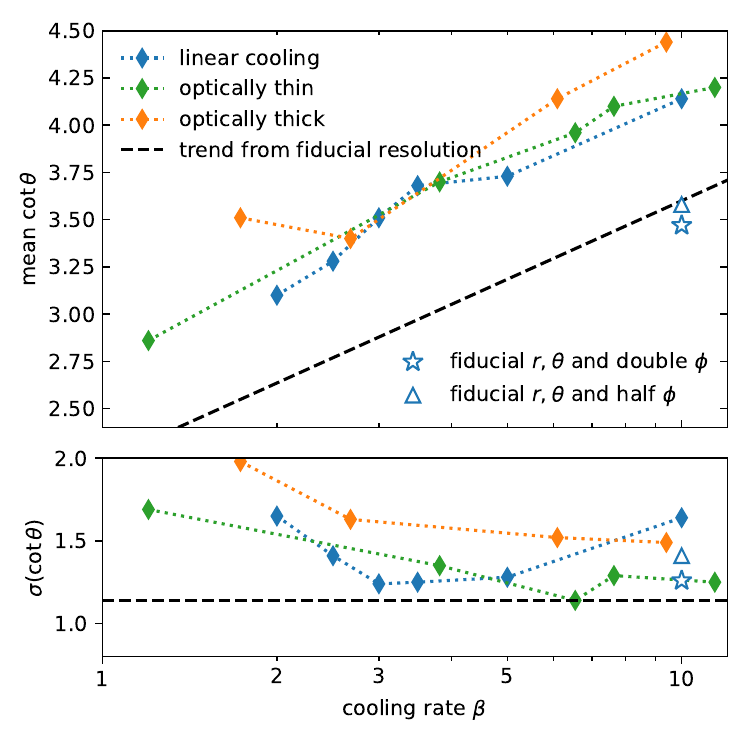}
    \caption{Similar to Fig.~\ref{fig:pitch_cooling_dependence}, but for the pitch-angle distribution of low-resolution runs. The $\beta$ dependence remains similar, but the pitch angle is more tightly wound and shows larger spread. This is probably associated with the higher numerical viscosity affecting clump dynamics. We also show runs at two other resolutions for $\beta=10$ linear cooling (open markers); these runs remain similar to the trend from fiducial-resolution runs, demonstrating converging behavior around fiducial resolution.}
    \label{fig:pitch_cooling_dependence_lowres}
\end{figure}

\begin{figure}
    \centering
    \includegraphics[scale=0.66]{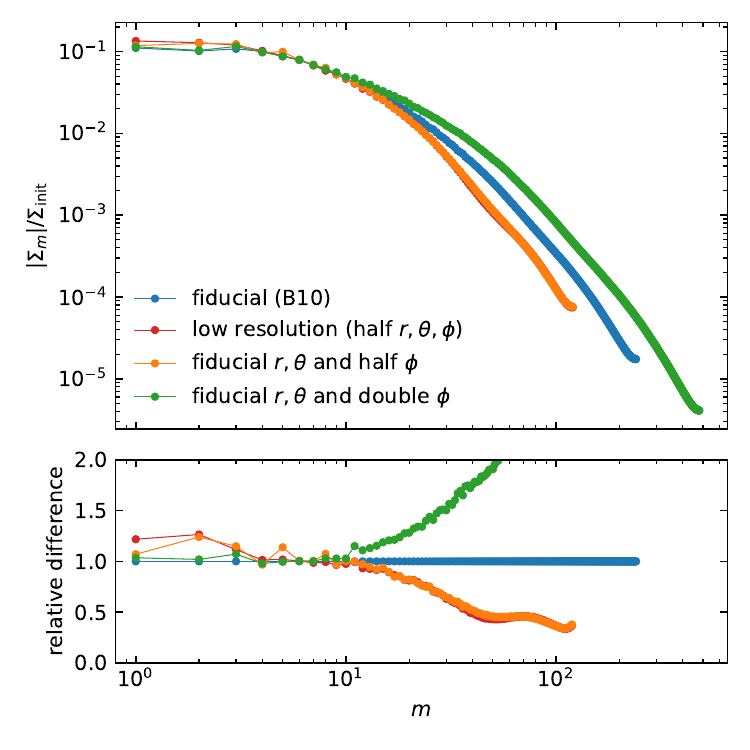}
    \caption{Spectrum for a $\beta=10$ disk at different resolutions. The bottom panel shows the relative difference compared to the fiducial resolution. As resolution increases, the amplitudes at low $m$, which captures most of the perturbation, show good agreement. At higher $m$, the spectrum is mainly controlled by $\phi$ resolution and is insensitive to the $r,\theta$ resolution, as shown by the agreement between the red and orange lines.}
    \label{fig:spectrum_res}
\end{figure}

Our low-resolution runs and fiducial-resolution runs agree with one another for most results featured in this paper, but one important exception is the distribution of pitch angles. The low-resolution runs have systematically more tightly wound spirals, as shown in Fig.~\ref{fig:pitch_cooling_dependence_lowres}. This may be due to spiral geometry being altered by the stronger numerical viscosity in low-resolution runs.
In Appendix~C of Paper~I, we argued that the numerical viscosity is likely strong enough to affect clump evolution in low-resolution runs but not in fiducial-resolution runs (where the numerical viscosity is ${\sim}8$ times weaker).
To test whether the pitch angle numerically converges around our fiducial resolution, we perform two new simulations at $\beta=10$ with fiducial $(r,\theta)$ resolution and half/double $\phi$ resolution.
(The $\phi$ resolution is likely the main source of numerical viscosity, because $\phi$ is the dimension with the lowest resolution and highest velocity.)
Both show pitch-angle distributions similar to those in the fiducial-resolution runs (empty markers in Fig.~\ref{fig:pitch_cooling_dependence_lowres}), suggesting that the pitch-angle distribution probably has reached numerical convergence around our fiducial resolution.

We comment that the exact reason why lower resolution produces more tightly wound spirals remains unclear. One hypothesis is that this is mainly due to different azimuthal ($m$) spectra, similar to how we explain the cooling rate dependence of pitch angle in Section~\ref{sec:obs:pitch}. However, comparing the azimuthal spectra across different resolutions disfavors this explanation. Fig.~\ref{fig:spectrum_res} shows that different resolutions show good agreement for $m\lesssim 10$, which dominates the perturbation amplitude. Additionally, the run with the fiducial $(r,\theta)$ resolution and half-$\phi$ resolution shows a spectrum that is almost identical to the low-resolution run, but the two show different pitch-angle distributions.  This suggests that the spectrum alone cannot explain the resolution dependence of pitch angle.

\bibliography{x24}{}
\bibliographystyle{aasjournal}

\end{CJK*}
\end{document}